\documentclass[a4paper,fleqn]{cas-dc}

\usepackage[numbers]{natbib}
\usepackage{float}
\usepackage{amsthm}
\usepackage{enumerate}
\newtheorem{theorem}{Proposition}

\makeatletter
\newcommand{\mathleft}{\@fleqntrue\@mathmargin0pt}
\newcommand{\mathcenter}{\@fleqnfalse}
\makeatother

\def\tsc#1{\csdef{#1}{\textsc{\lowercase{#1}}\xspace}}
\tsc{WGM}
\tsc{QE}
\tsc{EP}
\tsc{PMS}
\tsc{BEC}
\tsc{DE}

\newcommand{\changecolor}[1]{#1}

\begin{document}
\let\WriteBookmarks\relax
\def\floatpagepagefraction{1}
\def\textpagefraction{.001}
\shorttitle{A generalized perturbative approach for the computation of nonlinear scattering problems}
\shortauthors{J. Itier et~al.}

\title [mode = title]{A generalized perturbative approach for the computation of nonlinear scattering problems}                      


\author[1]{Jérémy Itier}[orcid=0009-0002-3783-1468]
\cormark[1]
\ead{jeremy.itier@fresnel.fr}

\affiliation[1]{organization={Aix Marseille Univ, CNRS, Centrale Med, Institut Fresnel},
                city={Marseille},
                country={France}}

\author[1]{Gilles Renversez}[orcid=0000-0002-7184-2665]
\author[1]{Frédéric Zolla}[orcid=0000-0003-1103-5041]

\cortext[cor1]{Corresponding author}

\begin{abstract}
We present a perturbative technique for modeling the scattering of light by a nonlinear material. This approach eliminates the need for an iterative algorithm to solve the fully coupled nonlinear problem. We demonstrate its effectiveness in the cases of a nonlinear anisotropic slab and a nonlinear periodic crystal, both illuminated by a plane wave under conical incidence and arbitrary polarization. Quantitative comparisons of the accuracy and computational time with a previously published rigorous model are provided.
\end{abstract}

\begin{graphicalabstract}
\includegraphics{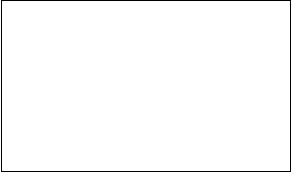}
\end{graphicalabstract}

\begin{highlights}
\item Nonlinear optics modeling
\item Perturbative approach
\item Finite element method
\item Scattering problem
\end{highlights}


\maketitle

\section{Introduction}
    Perturbation theory has long been a widely used tool in many fields, such as mechanics and chemistry. In particular, it is employed to describe complex quantum-mechanical systems in terms of simpler ones, but it has been less frequently applied in optics, where most problems were linear. Nevertheless, some applications have emerged, such as the study of small geometrical perturbations in waveguides~\cite{snyder_optical_1984, johnson_perturbation_2002}. 
    On the other hand, since the invention of lasers in the early \(1960s\) by Theodore Maiman, the power of light sources has steadily increased, revealing the nonlinear nature of light-matter interactions. From a modeling perspective, the nonlinearity of the constitutive relations leads, even for a monochromatic source, to a system of coupled nonlinear partial differential equations of a vector or even tensor nature~\cite{bloembergen_nonlinear_1996, zolla_into_2022}. 
    To solve these equations, two main approaches are commonly employed. The first, referred to as the \textit{rigorous approach} in the remainder of this article, consists in directly solving the coupled nonlinear system through iterative methods. Algorithms such as Picard or Newton-Raphson, combined with a linear equation solver (FD-TD~\cite{szarvas_numerical_2018}, finite element method~\cite{godard_optique_2009,walasik_plasmon-soliton_2016, elsawy_study_2017,elsawy_exact_2018,itier_scattering_2025},  etc.), can be used to compute the solution. However, this approach is time-consuming, typically requiring ten iterations or more even in favorable cases. In applications that are already computationally demanding, such as brute-force optimization, this method becomes extremely costly.
    A second approach relies on perturbation theory. Assuming that the field amplitudes are sufficiently small, the coupled nonlinear system can be decoupled into a set of linear equations that are solved sequentially. In the case of second-harmonic generation, this method is often applied in its simplest form by retaining only the first two equations, under the so-called undepleted-pump approximation~\cite{reinisch_electromagnetic_1983, reinisch_coupled-mode_1994, nakagawa_analysis_2002}. This approach has been used in Ref.~\cite{boyd_nonlinear_2020} for rough estimations of nonlinear susceptibilities, and its direct application for scattering problems has been reported in Ref.~\cite{zolla_into_2022-1} up to the third order. The extension of the perturbative method offers greater accuracy than the crude undepleted-pump approximation while remaining significantly faster than the rigorous approach. Moreover, it can be used to derive approximate analytical solutions.
    
    In this work, we generalize the latter approach to the \(n\)-th order for the full system of nonlinear equations describing light scattering in the harmonic domain and compare it, in several test cases, with the solution obtained from the rigorous approach reported in reference~\cite{itier_scattering_2025}. The configurations consist of a two-dimensional nonlinear slab and a nonlinear periodic crystal, studied for arbitrary polarizations and incident angles, and involving materials described by general susceptibility tensors. In both approaches, the nonlinear and linear equations are solved using the finite element method.
    
    The article is organized as follows. First, we recall the general formalism of the system of equations describing the vector electric field when second- and third-order nonlinear processes are considered. Second, using perturbation theory, we derive the general form of the perturbative equations. Third, we compare the two approaches in the case of second-harmonic generation in potassium titanyl phosphate (KTP) structures, providing quantitative error estimates as well as computation-time benchmarks. Fourth, we extend the study to second- and third-order effects in a lithium niobate (\(\mathrm{LiNbO_3}\)) nonlinear slab.

\section{Nonlinear scattering problem}
    Assuming a monochromatic incident field at the frequency \(\omega_I\), we suppose that the total field inside the nonlinear medium can be expanded as \(\mathbf{E}(\mathbf{r},t) = \sum_{p\in\mathbb{Z}^{*}}\mathbf{E}_p(\mathbf{r}) e^{-ip\omega_It}\), with \(\mathbf{E}_p\) and \(\mathbf{E}_{-p}\) being the complex amplitudes of the field and its conjugate both at frequency \(p\omega_I\). To test our method and to describe its capabilities, we focus only on second- and third-order nonlinearities, knowing that it can also tackle higher orders. Using the framework given in reference~\cite{zolla_into_2022}, the set of equations describing the scattering of light when considering only second- and third-order nonlinearities is:
    \begin{equation}
        \begin{array}{c}
             \mathbf{M}_p^{\mathrm{lin}}\mathbf{E}_p + \frac{\left(p\omega_I\right)^2}{c^2} \sum_{q\in \mathbb{Z}} \langle\langle  \mathbf{E}_{q}, \mathbf{E}_{p-q} \rangle\rangle \\
             + \frac{\left(p\omega_I\right)^2}{c^2} \sum_{(q,r) \in \mathbb{Z}^2} \langle\langle  \mathbf{E}_{q}, \mathbf{E}_{r}, \mathbf{E}_{p-q-r} \rangle\rangle = -i p \omega \mu_0 \,\mathbf{J}_{p} \delta_{|p|,1}
        \end{array}
        \label{eq:sys_nl}
    \end{equation}
    The linear part of the Maxwell operator is written \(\mathbf{M}_p^{\mathrm{lin}}\) and its definition is given in Eq.~\eqref{eq:Mlin}. The source term is represented by \(\mathbf{J}_p\delta_{|p|,1}\), where \(\delta_{i,j}\) is the Kronecker delta.
    The nonlinear parts of the equation are represented using the \(\langle\langle ... \rangle\rangle\) multi-linear operator, involving the susceptibility tensors \(\chi_{(n)}\) of the medium with the field amplitudes by means of contraction (\(:_{(n)}\)) and tensor product (\(\otimes\)), as shown in Eqs.~\eqref{eq:nl_operators}. 
    
    \begin{equation}
        \mathbf{M}_p^{\mathrm{lin}}\mathbf{E}_p  \equiv - \nabla \times \nabla \times \mathbf{E}_p + \frac{(p\omega_I)^2}{c^2}\varepsilon_r(p\omega_I) \mathbf{E}_p
        \label{eq:Mlin}
    \end{equation}
    \begin{equation}
        \begin{array}{lll}
            \varepsilon_r(p\omega_I) &\equiv 1 + \chi_{(1)}(p\omega_I)\\
            \langle\langle  \mathbf{E}_{p}, \mathbf{E}_{q} \rangle\rangle &\equiv \chi_{(2)}(p\omega_I, q\omega_I) : \mathbf{E}_p \otimes \mathbf{E}_q \\
            \langle\langle  \mathbf{E}_{p}, \mathbf{E}_{q}, \mathbf{E}_{r} \rangle\rangle &\equiv \chi_{(3)}(p\omega_I, q\omega_I, r\omega_I) :_{(3)} \mathbf{E}_p \otimes \mathbf{E}_q \otimes \mathbf{E}_r
        \end{array}
        \label{eq:nl_operators}
    \end{equation}
    
    The nonlinear system formally involves an infinite set of equations indexed by \(p\in\mathbb{Z}^*\). Since half of these equations are the complex conjugates of the others, the analysis can be restricted to \(p\in\mathbb{N}^*\). In practice, high-order harmonics are negligible, and the system is therefore truncated to a finite number \(p_{max}\) of equations. Nevertheless, the resulting system remains a fully coupled nonlinear set of equations, which is challenging to solve numerically.
    
\section{Perturbative approach}
    \subsection{Derivation of the perturbative equations}
        We assume that the field amplitudes are relatively small, meaning that the nonlinear terms \(\langle\langle  \mathbf{E}_{p}, \mathbf{E}_{q} \rangle\rangle\) and \(\langle\langle  \mathbf{E}_{p}, \mathbf{E}_{q}, \mathbf{E}_{r} \rangle\rangle\) are small compared to \(\varepsilon_r\, \mathbf{E}_p\), or, at most, of the same order of magnitude. Introducing a perturbation parameter \(\eta\), we define \(\mathbf{J}_1 = \eta \mathbf{J}_1^{(1)}\) and approximate each field \(\mathbf{E}_p\) by a finite expansion in powers of \(\eta\):
        \begin{equation}
            \mathbf{E}_p = \sum_{j=1}^{n}\eta^{j} \, \mathbf{E}_p^{(j)}
            \label{eq:expansion}
        \end{equation}
        
        We then substitute Eq.~\eqref{eq:expansion} into Eq.~\eqref{eq:sys_nl} to obtain \(p_{max}\) polynomial equations of the form \(Q_p(\eta)=\sum_{j=1}^{n}c_{p,j}\,\eta^j=0\), as shown in~Eq.~\eqref{eq:sys_prtb}.
        \begin{align}
           &\sum_{j=1}^n \eta^j \mathbf{M}_p^{\mathrm{lin}}\mathbf{E}_p^{(j)} + \frac{\left(p\omega_I\right)^2}{c^2} \sum_{k,l=1}^n \eta^{l+k} \sum_{q\in \mathbb{Z}} \langle\langle  \mathbf{E}_{q}^{(l)}, \mathbf{E}_{p-q}^{(k)} \rangle\rangle \label{eq:sys_prtb} \nonumber\\
            &+ \frac{\left(p\omega_I\right)^2}{c^2} \sum_{k,l,m=1}^n \eta^{l+m+k} \sum_{(q,r)\in \mathbb{Z}^2} \langle\langle  \mathbf{E}_{q}^{(l)}, \mathbf{E}_{r}^{(m)}, \mathbf{E}_{p-q-r}^{(k)} \rangle\rangle \nonumber\\
            &= -i p \omega \mu_0 \, \eta \mathbf{J}_{p}^{(1)} \delta_{|p|,1}
        \end{align}
        
        For the polynomials to vanish, each coefficient \(c_{p,j}\) must be zero, leading to a total of \(n \times p_{max}\) equations.

        \begin{subequations}
        The set of equations \(c_{p,1}=0\) leads to the classical Maxwell linear equation shown in Eq.~\eqref{eq:eta_1}.
        \begin{equation}
            \mathbf{M}_p^{\mathrm{lin}}\mathbf{E}_p^{(1)} = -i p \omega \mu_0 \,\mathbf{J}_{1}^{(1)} \delta_{|p|,1}
            \label{eq:eta_1}
        \end{equation}
        
        For \(j > 1\), the systems resulting from \(c_{p,j}=0\) are:
        \begin{align}
            &\mathbf{M}_p^{\mathrm{lin}}\mathbf{E}_p^{(j)} + \frac{\left(p\omega_I\right)^2}{c^2} \sum_{l+k=j} \sum_{q\in \mathbb{Z}} \langle\langle  \mathbf{E}_{q}^{(l)}, \mathbf{E}_{p-q}^{(k)} \rangle\rangle \nonumber\\
            &+ \frac{\left(p\omega_I\right)^2}{c^2} \sum_{l+m+k=j} \sum_{(q,r)\in \mathbb{Z}^2} \langle\langle  \mathbf{E}_{q}^{(l)}, \mathbf{E}_{r}^{(m)}, \mathbf{E}_{p-q-r}^{(k)} \rangle\rangle = \mathbf{0}
            \label{eta_j}
        \end{align} \label{eq:sys_prtb_2}
        \end{subequations}
        
        At first sight, sys.~\eqref{eq:sys_prtb_2} may appear demanding, but we show in the next two sections that most of its terms vanish, which drastically reduces the number of equations that must be solved. Most importantly, the system takes the form of a cascade system as shown in Fig.~\ref{fig:triangle_prtb}: once a field \(\mathbf{E}_p^{(j)}\) has been computed for given \(j\) and \(p\), it becomes a known nonlinear source term for the subsequent linear problems.
        
    \subsection{General properties}
        First, surprisingly enough, the resulting fields \(\mathbf{E}_p^{(j)}\) are independent of the value of the small parameter \(\eta\), which allows us to set \(\eta=1\) in the simulations.
        
        Second, the perturbative field \(\mathbf{E}_p^{(j)}\) can be characterized by two properties that define the conditions under which it vanishes:
        \begin{enumerate}[label=(\roman*)]
            \item If \(p\) is higher than \(j\), then the corresponding field is zero. \\
            \(\forall (p,j) \in \mathbb{Z}\times \mathbb{N},\, |p| > j \implies \mathbf{E}_p^{(j)}=\mathbf{0}\)
            \item If \(p\) and \(j\) do not have the same parity, then the corresponding field is zero. \\
            \( \forall p \in \mathbb{Z},\, p-j = 1[2] \implies \mathbf{E}_p^{(j)}=\mathbf{0}\)
        \end{enumerate}
        The proofs are provided in the appendix.

\begin{figure}
            \centering
            \includegraphics[width=0.7\linewidth]{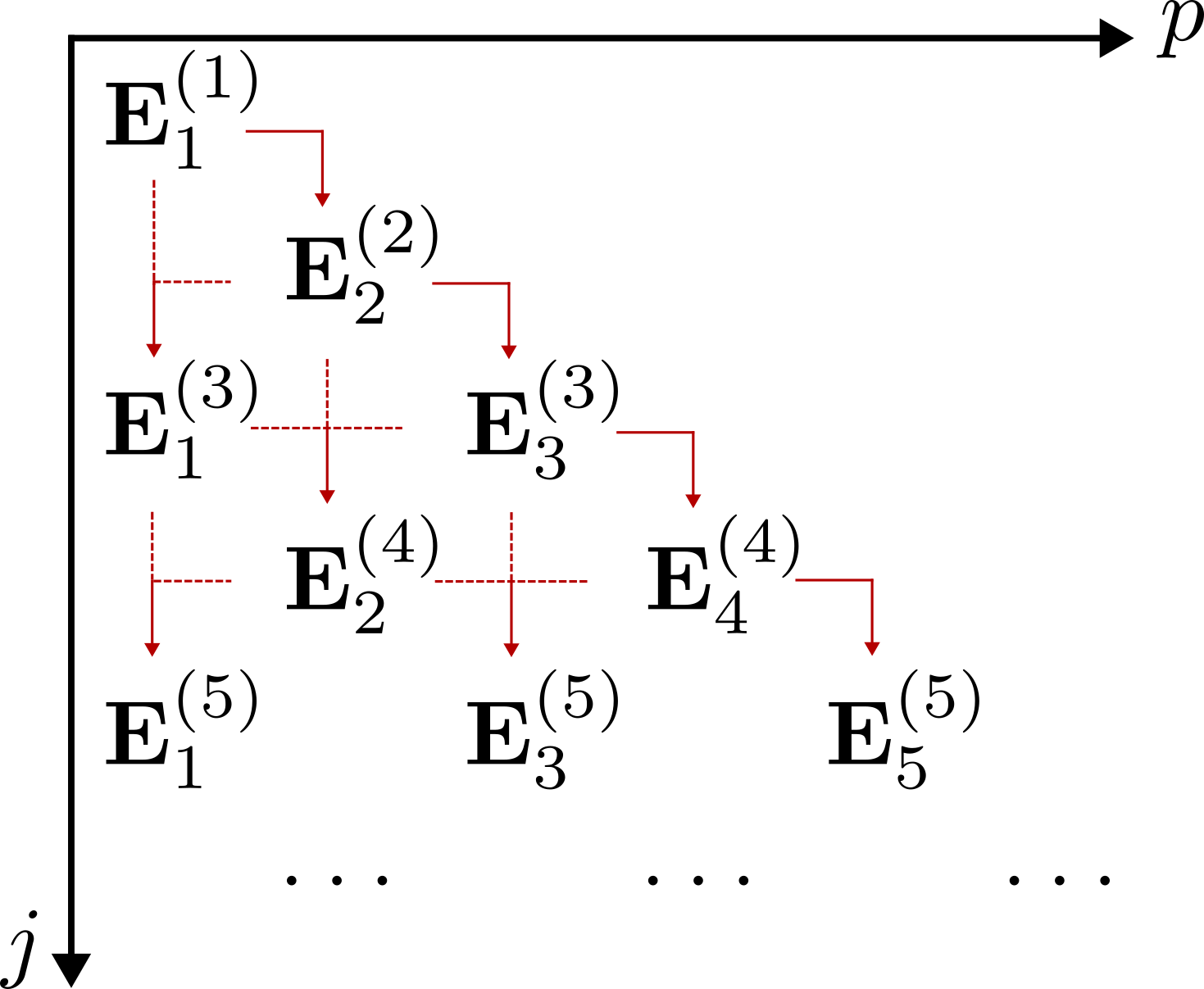}
            \caption{Illustration of the cascading triangular system. The indices \(j\) and \(p\) denote the perturbative order and the harmonic number, respectively. Each pointed field \(\mathbf{E}_p^{(j)}\) depends only on the connected elements.}
            \label{fig:triangle_prtb}
        \end{figure}

        Using the two latter properties, we obtain a triangular system of equations, as illustrated in Fig.~\ref{fig:triangle_prtb}. As with a standard triangular matrix system, each equation can be solved sequentially using the solutions computed at previous steps. 

    \subsection{Explicit derivation}
        We start by developing the summations indexed on the perturbative orders \(j\), following the tables~\ref{tab:combinaison_1} and~\ref{tab:combinaison_2}.
        The first five systems corresponding to \(c_{p,1}=0\), ..., \(c_{p,5}=0\) are shown in Sys.~\eqref{eq:sys_prtb_j}.
        \begin{subequations}
            \mathleft
            \begin{align}
                \mathbf{M}_p^{\mathrm{lin}}\mathbf{E}_p^{(1)}& = -i p \omega \mu_0 \,\mathbf{J}_{p}^{(1)} \delta_{|p|,1} \\
                \mathbf{M}_p^{\mathrm{lin}}\mathbf{E}_p^{(2)}& + \frac{\left(p\omega_I\right)^2}{c^2} \sum_{q\in \mathbb{Z}} \left( \langle\langle  \mathbf{E}_{q}^{(1)}, \mathbf{E}_{p-q}^{(1)} \rangle\rangle \right) = \mathbf{0} \\
                \mathbf{M}_p^{\mathrm{lin}}\mathbf{E}_p^{(3)}& + \frac{\left(p\omega_I\right)^2}{c^2} \sum_{q\in \mathbb{Z}} \left( 2\langle\langle  \mathbf{E}_{q}^{(1)}, \mathbf{E}_{p-q}^{(2)} \rangle\rangle \right) \nonumber\\[-5pt]
                & + \frac{\left(p\omega_I\right)^2}{c^2} \sum_{(q,r)\in \mathbb{Z}^2} \left( \langle\langle  \mathbf{E}_{q}^{(1)}, \mathbf{E}_{r}^{(1)}, \mathbf{E}_{p-q-r}^{(1)} \rangle\rangle \right) = \mathbf{0} \\
                \mathbf{M}_p^{\mathrm{lin}}\mathbf{E}_p^{(4)}& + \frac{\left(p\omega_I\right)^2}{c^2} \sum_{q\in \mathbb{Z}} \left( 2\langle\langle \mathbf{E}_{q}^{(1)}, \mathbf{E}_{p-q}^{(3)} \rangle\rangle + \langle\langle  \mathbf{E}_{q}^{(2)}, \mathbf{E}_{p-q}^{(2)} \rangle\rangle \right) \nonumber\\[-5pt]
                &+ \frac{\left(p\omega_I\right)^2}{c^2} \sum_{(q,r)\in \mathbb{Z}^2} \left( 3\langle\langle  \mathbf{E}_{q}^{(2)}, \mathbf{E}_{r}^{(1)}, \mathbf{E}_{p-q-r}^{(1)} \rangle\rangle \right) = \mathbf{0} \\
                \mathbf{M}_p^{\mathrm{lin}}\mathbf{E}_p^{(5)}& + \frac{\left(p\omega_I\right)^2}{c^2} \sum_{q\in \mathbb{Z}} \left( 2\langle\langle \mathbf{E}_{q}^{(1)}, \mathbf{E}_{p-q}^{(4)} \rangle\rangle + 2\langle\langle  \mathbf{E}_{q}^{(2)}, \mathbf{E}_{p-q}^{(3)} \rangle\rangle \right) \nonumber\\[-5pt]
                &+ \frac{\left(p\omega_I\right)^2}{c^2} \sum_{(q,r)\in \mathbb{Z}^2} \Bigl( 3\langle\langle  \mathbf{E}_{q}^{(3)}, \mathbf{E}_{r}^{(1)}, \mathbf{E}_{p-q-r}^{(1)} \rangle\rangle  \nonumber\\
                &+ 3\langle\langle  \mathbf{E}_{q}^{(2)}, \mathbf{E}_{r}^{(2)}, \mathbf{E}_{p-q-r}^{(1)} \rangle\rangle \Bigr) = \mathbf{0}
            \end{align} \label{eq:sys_prtb_j}
            \mathcenter
        \end{subequations}

        \begin{table}[H]
            \centering
            \caption{\bf Combinations \(\mathbf{l+k=j}\)}
            \begin{tabular}{c|c|cc|ccc|cccc}
                j & 2 & \multicolumn{2}{c|}{3} & \multicolumn{3}{c|}{4} & \multicolumn{4}{c}{5} \\ \hline
                l & 1 & 1          & 2         & 1      & 2     & 3     & 1   & 2   & 3   & 4   \\
                k & 1 & 2          & 1         & 3      & 2     & 1     & 4   & 3   & 2   & 1  
            \end{tabular}
            \label{tab:combinaison_1}
        \end{table}
        \begin{table}[H]
            \centering
            \caption{\bf Combinations \(\mathbf{l+k+m=j}\)}
            \begin{tabular}{c|c|ccc|cccccc}
                j & 3 & \multicolumn{3}{c|}{4} & \multicolumn{6}{c}{5} \\ \hline
                l & 1 & 1 & 1 & 2 & 1 & 1 & 1 & 2 & 2 & 3 \\
                k & 1 & 1 & 2 & 1 & 1 & 2 & 3 & 1 & 2 & 1 \\
                \multicolumn{1}{l|}{m} & \multicolumn{1}{l|}{1} & \multicolumn{1}{l}{2} & \multicolumn{1}{l}{1} & \multicolumn{1}{l|}{1} & \multicolumn{1}{l}{3} & \multicolumn{1}{l}{2} & \multicolumn{1}{l}{1} & \multicolumn{1}{l}{2} & \multicolumn{1}{l}{1} & \multicolumn{1}{l}{1}
            \end{tabular}
            \label{tab:combinaison_2}
        \end{table}
        
        Using properties (i) and (ii) of Sec.3.2, we can further reduce the system of equations:
        \begin{itemize}
            \item if \(j\) is odd, only the fields at \(p\omega_I\) with \(p=1, 3, 5, ..., j\) are nonzero;
            \item if \(j\) is even, only the fields at \(p\omega_I\) with \(p=2, 4, 6, ..., j\) are nonzero.
        \end{itemize}
        
        We then expand the final summation over the harmonics \(p\), as shown concisely in Eq.~\eqref{eq:sys_prtb_explicit}, to obtain the set of equations describing the perturbative approach. The explicit expressions of the source terms, denoted \(\mathbf{S}_p^{(j)}\), are provided in Appendix B up to the fifth order. 
        \begin{subequations}
            \begin{align}
                \mathbf{M}_1^{\mathrm{lin}}\mathbf{E}_1^{(1)}& = -i \omega \mu_0 \,\mathbf{J}_{p}^{(1)} \label{eq:sys_prtb_explicit_1} \\
                \mathbf{M}_p^{\mathrm{lin}}\mathbf{E}_p^{(j)}& + \mathbf{S}_p^{(j)} \bigl(\mathbf{E}_q^{(l)}\bigr) = \mathbf{0}
            \end{align} \label{eq:sys_prtb_explicit}
        \end{subequations}
        with \(q<p\) and \(l<j\).

        Since each nonlinear source \(\mathbf{S}_p^{(j)}\) depends only on previously computed fields, Sys.~\eqref{eq:sys_prtb_explicit} can be solved sequentially using a linear solver. Starting from Eq.~\eqref{eq:sys_prtb_explicit_1}, each linear equation is solved using the previously obtained fields as input variables in the nonlinear source terms \(\mathbf{S}_p^{(j)}\). The fields \(\mathbf{E}_p\) are then recovered using Eq.~\eqref{eq:expansion}, i.e., \(\mathbf{E}_p=\sum_j \mathbf{E}_p^{(j)}\). We note that at order \(j\), only harmonics up to the \(j\)th are accounted for; modeling higher-order harmonics would require extending the perturbative expansion further, as a direct consequence of property (i).

\section{Application: Second-harmonic generation}
    \subsection{Governing equations}
        To obtain the system of equations in the case of second-harmonic generation (2HG), we discard the third-order nonlinearities and retain only the first two equations of Sys.~\eqref{eq:sys_nl}.
        We end up with the usual second-harmonic generation system seen in the literature~\cite{bloembergen_nonlinear_1996}, as shown in \eqref{eq:2HG}.
        \begin{subequations}
            \begin{align}
                &\mathbf{M}_1^{\mathrm{lin}}\mathbf{E}_1 + 2\frac{\omega_I^2}{c^2} \, \langle\langle  \mathbf{E}_{-1}, \mathbf{E}_2 \rangle\rangle = -i\omega\mu_0\,\mathbf{J}_{1} \label{eq:2HG_1}\\
                &\mathbf{M}_2^{\mathrm{lin}}\mathbf{E}_2 + \frac{\left(2\omega_I\right)^2}{c^2} \, \langle\langle  \mathbf{E}_1, \mathbf{E}_1 \rangle\rangle = \mathbf{0}
                \label{eq:2HG_2}
            \end{align} \label{eq:2HG}
        \end{subequations}

        This set of equations forms a fully coupled nonlinear system. By that, we mean that Eq.~\eqref{eq:2HG_1} requires \(\mathbf{E}_2\) to obtain \(\mathbf{E}_1\) and conversely Eq.~\eqref{eq:2HG_2} requires \(\mathbf{E}_1\) to obtain \(\mathbf{E}_2\). The purpose of the perturbative approach is precisely to decouple this system.
        
        By considering the system \eqref{eq:sys_prtb_explicit} with \(p_{max}=2\) and \(j_{max}=5\), we obtain the set of equations presented in Sys.~\eqref{eq:sys_prtb_2HG}.
        \begin{subequations}
            \begin{align}
                \mathbf{M}_1^{\mathrm{lin}}\mathbf{E}_1^{(1)}& = -i \omega \mu_0 \,\mathbf{J}_{1}^{(1)} \label{eq:sys_prtb_2HG_1}\\
                \mathbf{M}_2^{\mathrm{lin}}\mathbf{E}_2^{(2)}& + \frac{(2\omega_I)^2}{c^2}  \langle\langle  \mathbf{E}_{1}^{(1)}, \mathbf{E}_{1}^{(1)} \rangle\rangle = \mathbf{0} \label{eq:sys_prtb_2HG_2}\\
                \mathbf{M}_1^{\mathrm{lin}}\mathbf{E}_1^{(3)}& + 2\frac{\omega_I^2}{c^2}  \langle\langle  \mathbf{E}_{-1}^{(1)}, \mathbf{E}_{2}^{(2)} \rangle\rangle = \mathbf{0} \label{eq:sys_prtb_2HG_3}\\
                \mathbf{M}_2^{\mathrm{lin}}\mathbf{E}_2^{(4)}& + 2\frac{(2\omega_I)^2}{c^2} \langle\langle  \mathbf{E}_{1}^{(1)}, \mathbf{E}_{1}^{(3)} \rangle\rangle = \mathbf{0} \label{eq:sys_prtb_2HG_4}\\
                \mathbf{M}_1^{\mathrm{lin}}\mathbf{E}_1^{(5)}& + 2\frac{\omega_I^2}{c^2}  \bigl(\langle\langle  \mathbf{E}_{-1}^{(1)}, \mathbf{E}_{2}^{(4)} \rangle\rangle + \langle\langle  \mathbf{E}_{-1}^{(3)}, \mathbf{E}_{2}^{(2)} \rangle\rangle\bigr) = \mathbf{0} \label{eq:sys_prtb_2HG_5}
            \end{align} \label{eq:sys_prtb_2HG}
        \end{subequations}
        
        Equations~\eqref{eq:sys_prtb_2HG_1} and~\eqref{eq:sys_prtb_2HG_2} correspond to the expressions commonly used in the literature under the undepleted-pump approximation (Eq.~7-10 of Ref.~\cite{nakagawa_analysis_2002}). By including the third equation~\eqref{eq:sys_prtb_2HG_3}, pump depletion effects can also be accounted for: the supplementary term $\mathbf{E}_1^{(3)}$ in the perturbative expansion of $\mathbf{E}_p$ provided by Eq.~\eqref{eq:expansion} is the first non-null correction term in the pump field.
        Adding Eqs.~\eqref{eq:sys_prtb_2HG_4} and~\eqref{eq:sys_prtb_2HG_5} is expected to further improve the accuracy of the resulting solution.
        
\begin{figure}
            \centering
            \fbox{\includegraphics[width=0.7\linewidth]{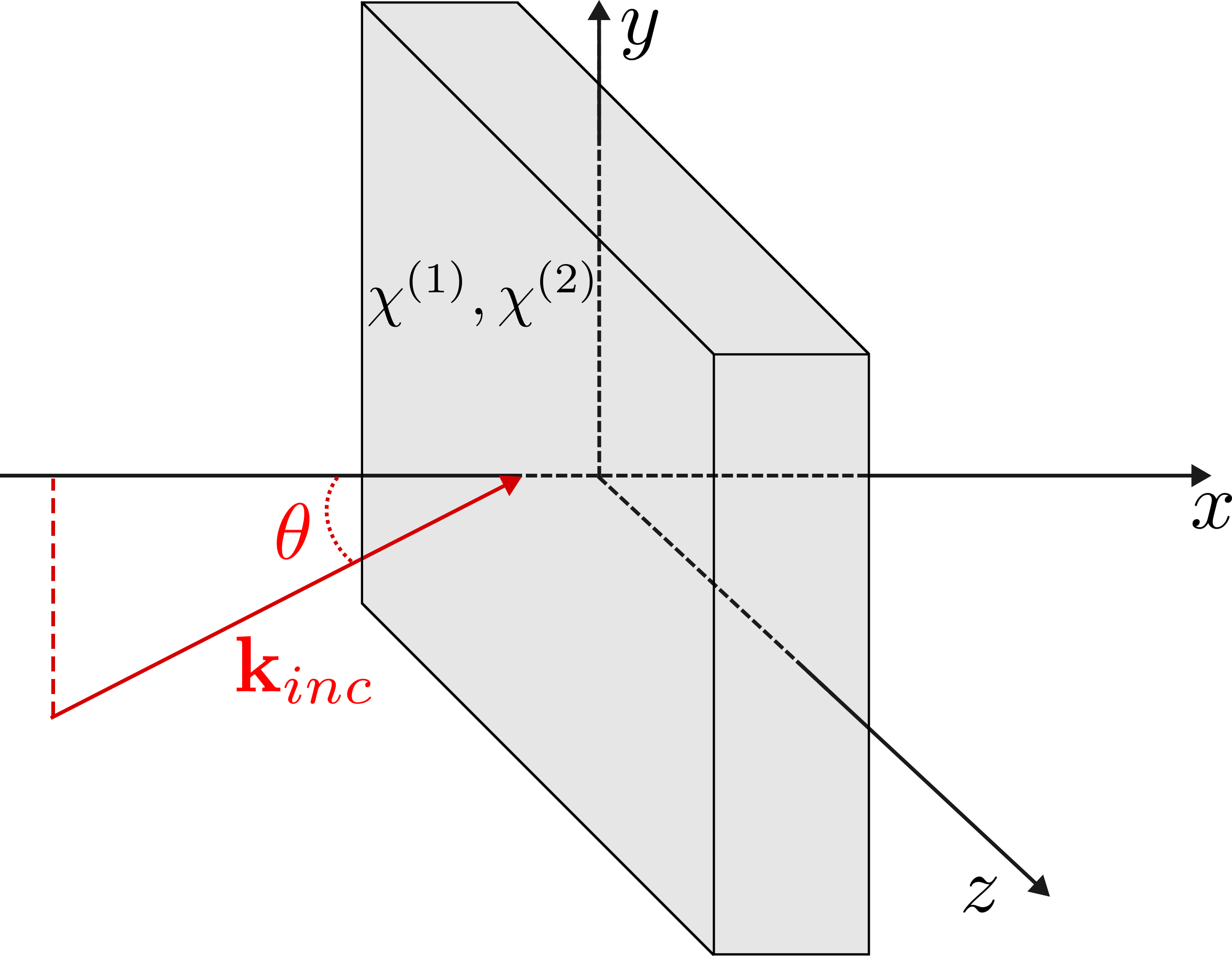}}
            \caption{Schematic view of a nonlinear slab illuminated by a plane wave with its wave vector \(\mathbf{k}_{inc}\) on its left side. \(\mathbf{k}_{inc}\) is within the \((x,y)\) plane.}
            \label{fig:schematic}
        \end{figure}

        To illustrate the approach, we consider the nonlinear slab depicted in Fig.~\ref{fig:schematic}, which is infinite along the \(y\)- and \(z\)-axes and characterized by the susceptibility tensors \(\chi^{(1)}\) and \(\chi^{(2)}\).
        The structure is illuminated by a plane wave under conical incidence. Nevertheless, we can choose a basis such that the wave vector \(\mathbf{k}_{inc}\) is contained in the \((\mathbf{e}_x, \mathbf{e}_y)\) plane. In return, the tensors \(\chi^{(1)}\) and \(\chi^{(2)}\) are a priori arbitrary. We note \(\theta\) the angle between the \(x\)-axis and \(\mathbf{k}_{inc}\). Let \(\mathbf{E}_{inc}\) be the complex amplitude of this incident wave:
        \begin{equation}
            \mathbf{E}_{inc}(x,y) =  \mathbf{A}_0 e^{i(\alpha x + \beta y)} = \mathbf{A}_0 e^{ik_0(\cos{\theta} \,x + \sin{\theta} \,y)}
            \label{eq:Einc}
        \end{equation}
        with \(\mathbf{A}_0\) and \(k_0\) being the amplitude and the wave number of the incident wave.
        
        We denote by \(T_y = \frac{2\pi}{k_0 \sin{\theta}}\) the periodicity of the incident wave along the \(y\)-direction. Since both the incident wave and the problem's geometry are \(T_y\)-periodic along \(\mathbf{e}_y\), the resulting field inside the slab must also exhibit \(T_y\)-periodicity. In the special case of normal incidence, the problem becomes independent of \(y\), and the field is accordingly uniform along that direction. Moreover, as shown in \cite{itier_scattering_2025}, a solution of the form \(\mathbf{E}_p^{(j)}(x,y) = \mathbf{\tilde{E}}_p^{(j)}(x) e^{ip\beta y}\) satisfies the governing equations. The problem thus reduces to a one-dimensional formulation, where the system given in Sys.~\eqref{eq:sys_prtb_2HG} remains valid provided that the linear operator \(\mathbf{M}^{\mathrm{\mathrm{lin}}}_p\) is replaced by \(\mathbf{\tilde{M}}^{\mathrm{lin}}_p\), defined as:
        \begin{equation}
           \mathbf{\tilde{M}}^{\mathrm{lin}}_p\left(\mathbf{\tilde{E}}_p^j(x)\right) \equiv  \mathbf{M}^{\mathrm{lin}}_p\left(\mathbf{\tilde{E}}_p^j(x) e^{ip\beta y}\right) e^{-ip\beta y}  
        \end{equation}
        Note that \(\mathbf{\tilde{M}}^{\mathrm{lin}}_p\left(\mathbf{\tilde{E}}_p^j(x)\right)\) only depends upon the \(x\) variable. An explicit expression is given in Appendix C.
    
    \subsection{Computation and energy balance}
        The FEM simulations were performed using \textit{Gmsh} \cite{geuzaine_gmsh_2009} and \textit{GetDP} \cite{dular_general_1998, geuzaine_getdp_2007}, two open-source software packages that have been widely used to model partial differential equations with non-trivial boundary conditions, including electromagnetic ones~\cite{drouart_spatial_2008, zolla_foundations_2012}. 
        A virtual antenna technique was employed to simulate the incident wave~\cite{zolla_virtual_2006}. Outgoing wave conditions were applied at the domain boundaries as they are exact in one-dimensional problems~\cite{schot_eighty_1992}.

        In order to assess the convergence of our numerical models, we developed a specific energy conservation rule from basic principles~\cite{itier_scattering_2025}. It allows us to quantify the energy exchange between the harmonics themselves and the material. In the case of a slab, it can be written:
        \begin{equation}
            \sum_p(R_p+T_p)+Q-1=0 \label{eq:energy_study}
        \end{equation}
        with \(R_p\) and \(T_p\) the reflection and transmission coefficients at frequency \(p\omega_I\), and \(Q\) the total electromagnetic losses inside the material, including both linear and nonlinear contributions at the different frequencies involved.
    
    \subsection{Numerical results}
        For clarity, we start our analysis with the TE-polarization case. We consider a KTP slab belonging to the \textit{mm2} point group symmetry. In the crystallographic basis in which the tensor coefficients are commonly defined, an incident TE wave remains TE as it propagates through the slab~\cite{itier_scattering_2025}. In this configuration, both the permittivity and the nonlinear susceptibility tensors can be reduced to single effective coefficients, \(n_e\) and \(\chi^{(2)}_{zzz}\). According to Ref.~\cite{fan_second_1987} and~\cite{shoji_absolute_1997}, the corresponding parameters are \(n_z^{\omega_I}=1.8302\), \(n_z^{2\omega_I}=1.8888\) and \(\chi^{(2)}_{zzz}=2d_{33}=29.2 \,.10^{-12}\,\mathrm{m/V}\). The  slab thickness is set to 2  $\mu$m.
        
\begin{figure*}
            \centering
            \includegraphics[width=0.8\linewidth]{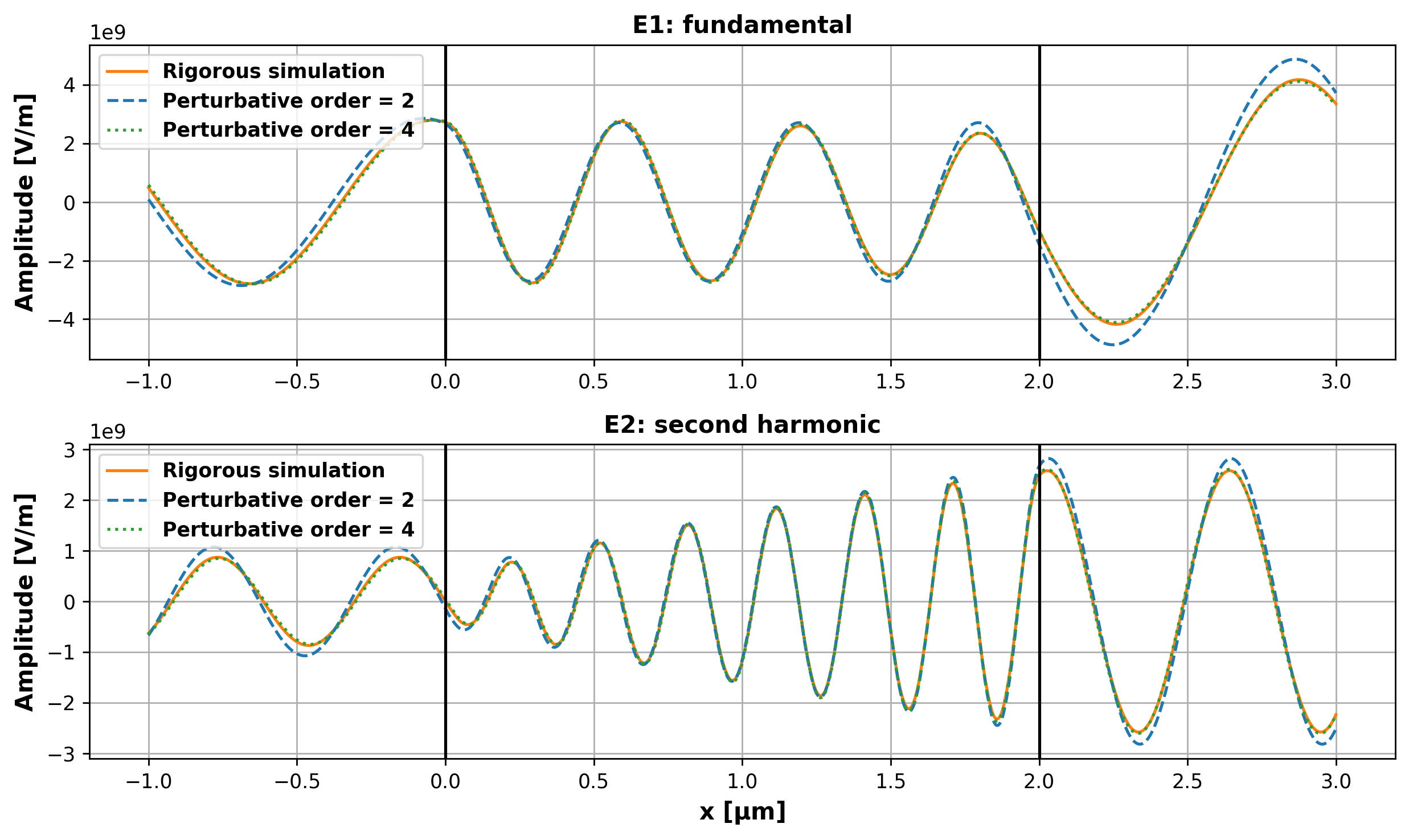}
            \caption{Nonlinear scattering from a KTP slab under TE-polarized illumination. The real parts of the fundamental field \(\mathbf{E}_1\) and the second-harmonic \(\mathbf{E}_2\) are plotted along the \(x\)-axis (\(y=0\)), for an incident plane wave from the left with amplitude \(A_0=6\times10^{9}V/m\) and incidence angle \(\theta=30^{\circ}\). The three curves are associated to the rigorous method, the second-order perturbative method, and the fourth-order perturbative method, respectively.} 
            \label{fig:2HG_field}
        \end{figure*}

        A plot of the electric field for an incident TE plane wave on the KTP slab is shown in Fig.~\ref{fig:2HG_field}. The field is computed in three ways: (i) the rigorous method, in which the nonlinear system is solved iteratively as reported in~\cite{itier_scattering_2025}; (ii) the second-order perturbative approach with \(j_{max}=2\); and (iii) the fourth-order perturbative approach with \(j_{max}=4\).
        An increasing second-harmonic is generated as the wave passes through the slab while the fundamental field slightly decreases.
        As expected, the second-order perturbative approach does not account for pump depletion, resulting in a noticeable discrepancy in the first and second harmonics. However, including two additional terms in the perturbative expansion improves the accuracy, yielding results that more closely match the rigorous solution. The corresponding energy balance errors are \(2\times10^{-6}\%\), \(25\%\) and \(1.9\%\) for the three simulations, respectively. This indicates that adding only a few terms in the perturbative expansion can drastically improve the simulation accuracy without resorting to a time-consuming iterative scheme.
        
\begin{figure}
            \centering
            \fbox{\includegraphics[width=0.9\linewidth]{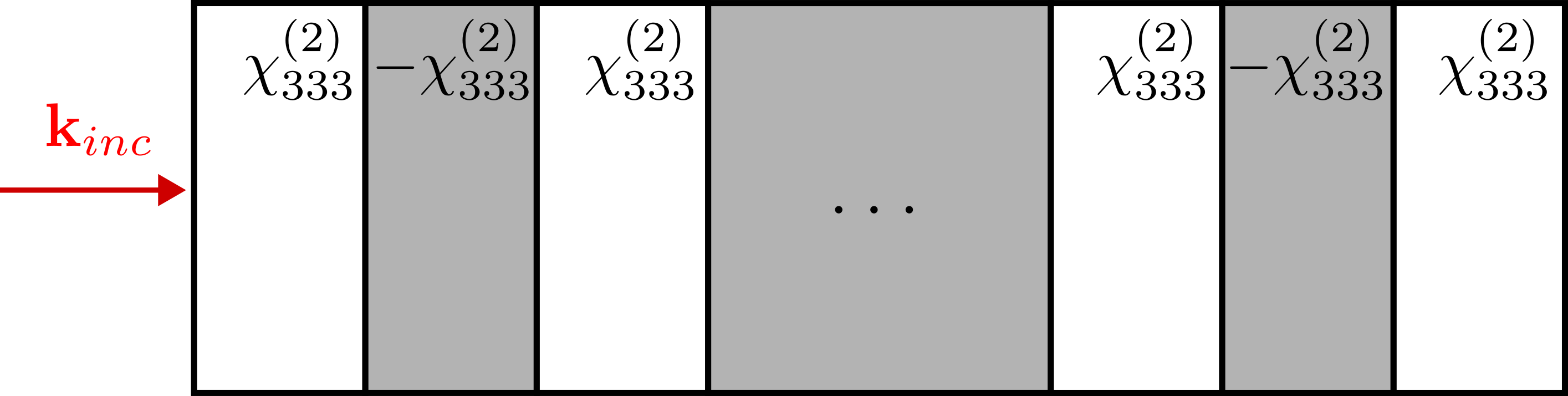}}
            \caption{\changecolor{Schematic view of a nonlinear periodic crystal illuminated by a plane wave in normal incidence.}}
            \label{fig:schematic_periodic_crystal}
        \end{figure}

        \smallskip
        In the next example, we reproduce the setup of Ref.~\cite{szarvas_numerical_2018}, Sec.~3.B, replacing the material with KTP. We consider the photonic crystal \changecolor{shown in Fig.~\ref{fig:schematic_periodic_crystal}} illuminated by a normally incident plane wave of amplitude \(A_0=4\times10^8\,\mathrm{V/m}\). The crystal consists of alternating layers of KTP, with one of the two layers rotated by 180° around the x-axis. As a result, both layers share the same refractive indices \(n_z^{\omega_I}\) and \(n_z^{2\omega_I}\), but their nonlinear coefficients \(\chi_{zzz}^{(2)}\) have opposite signs. The length \(l\) of the layer is set to be equal to the coherence length of the second-harmonic generation, with \(l=\frac{\lambda_0}{4|n_e^{2\omega}-n_e^{\omega}|}=4.539\,\mathrm{\mu m}\).
        Index matching is enforced at both ends of the crystal to avoid reflections. Without reflections, backward-propagating waves are negligible compared to forward-propagating ones, so the transmission coefficients can be regarded as the conversion efficiency.
        \begin{figure}
            \centering
            \includegraphics[width=\linewidth]{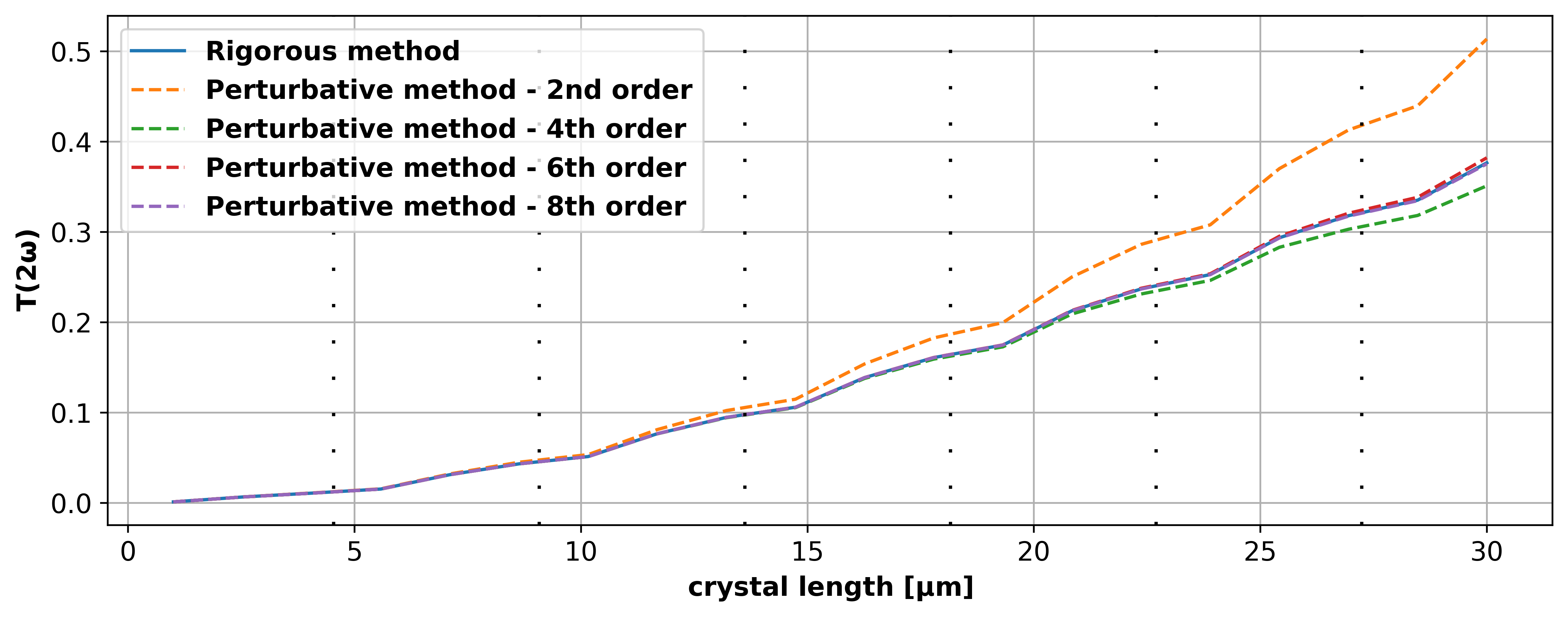}
            \caption{Transmission coefficient \(T_2\) of the second harmonic as a function of the photonic crystal length, computed using both the rigorous and perturbative methods. The simulations were performed for a TE-polarized plane wave at normal incidence. The dotted vertical bars indicate the positions of the layer interfaces. }
            \label{fig:Transmission2HG_photonic_cristal}
        \end{figure}
        
        Fig.~\ref{fig:Transmission2HG_photonic_cristal} shows the transmission coefficient \(T_2\) of the second harmonic as a function of the crystal length, computed using both the rigorous and perturbative methods. In agreement with expectations, the undepleted-pump approximation largely overestimates the energy conversion, with a deviation of \(36.39\%\) for a \(30\,\mathrm{\mu m}\)-thick crystal. In contrast, higher perturbative orders are significantly more accurate, with relative errors of \(6.76\%\), \(1.48\%\) and \(0.31\%\) for perturbative orders 4, 6 and 8, respectively. This demonstrates that even for relatively high conversion efficiency \((38\%)\), the perturbative method provides an accurate approximation of the rigorous model.

        \smallskip
    \subsection{Convergence of the method}
    
        To quantitatively assess the accuracy of the perturbative approach and investigate the convergence of the perturbative field as the expansion order increases, we compute the relative error \(e_r\) between the approximated field obtained from the perturbative approach and the one from the rigorous solution, defined as:
        \begin{equation}
            e_r = \frac{||(E_p)_z^{\mathrm{rigorous}}-(E_p)_z^{\mathrm{perturbatif}}||_2}{||(E_p)_z^{\mathrm{rigorous}}||_2}
        \end{equation}
        with \(||\cdot,\cdot||_2\) the euclidean norm:
        \begin{equation}
            ||A||_2 = \sqrt{\int |A(x, 0)|^2\,\mathrm{d}x} 
        \end{equation}   
        
\begin{figure}
            \centering
            \includegraphics[width=\linewidth]{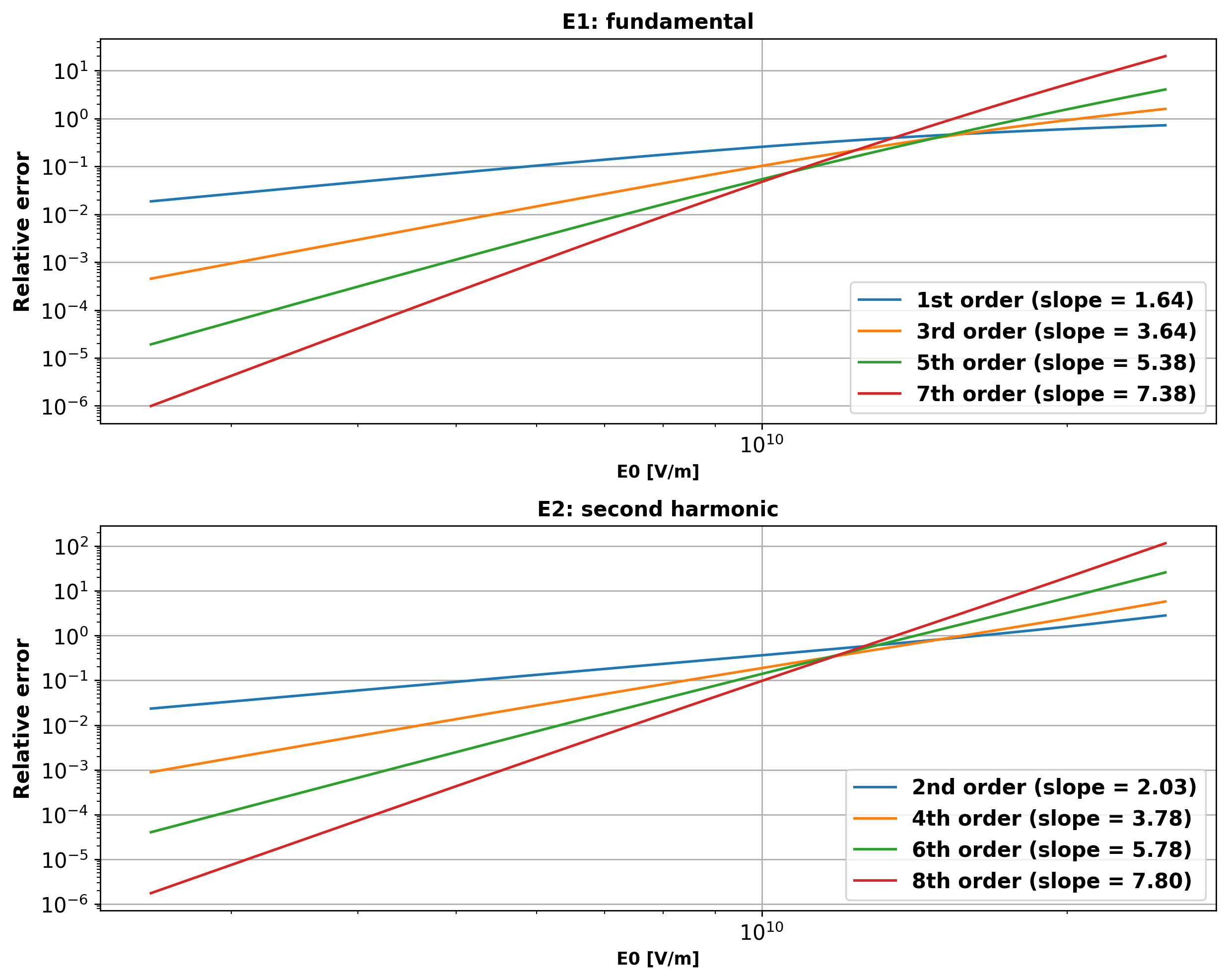}
            \caption{Relative errors between the complex field amplitudes obtained with the rigorous and perturbative approaches as a function of the incident-wave amplitude, for various expansion orders, plotted on a logarithmic scale.}
            \label{fig:2HG_conv}
        \end{figure}

        Fig.~\ref{fig:2HG_conv} shows the relative error between the complex field amplitudes along the \(x\)-axis obtained with the rigorous and perturbative approaches. The study is conducted for a \(2\,\mathrm{\mu m}\) KTP slab illuminated by a TE-polarized plane wave at an incident angle \(\theta=30^{\circ}\).            
        Since in the case of second-harmonic generation increasing the perturbative order alternately improves either the fundamental or the second harmonic, as shown in Eq.~\ref{eq:sys_prtb_2HG}, only the final odd-order perturbative expansions are plotted for the fundamental, and only final even-order perturbative expansions for the second harmonic, in order to avoid overloading the figures.
        
        We observe that the error decreases approximately with the power of the perturbative order: \(e_r\sim (E_0)^j\). At low energy, increasing the perturbative order greatly improves the accuracy of the model. However, the perturbative expansion has a finite radius of convergence; beyond this limit, the series no longer converges, and further increasing the order only amplifies the error.

\begin{figure}
            \centering
            \includegraphics[width=\linewidth]{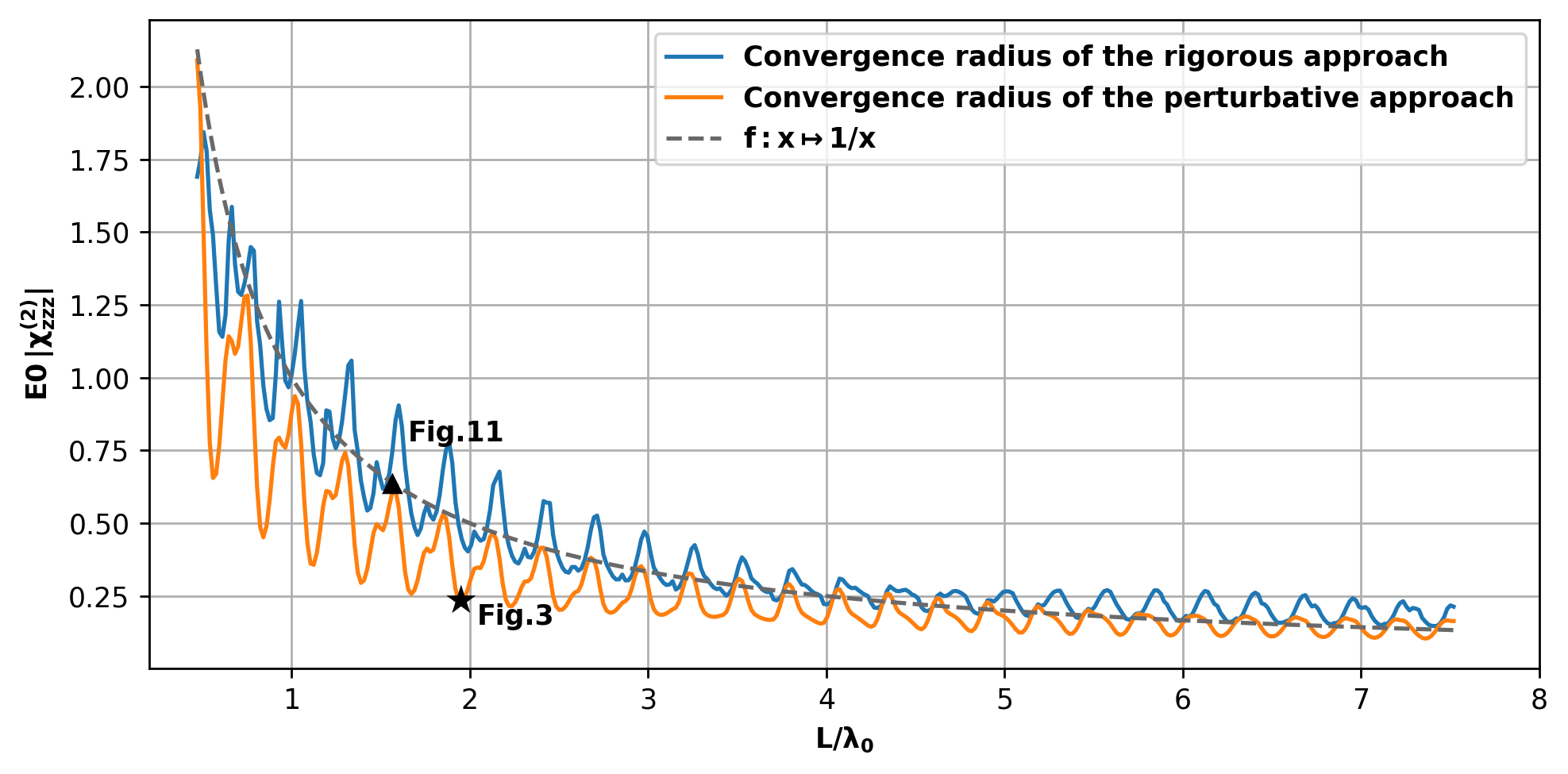}
            \caption{\changecolor{Convergence radius plotted in the \((L/\lambda_0,\ |\chi^{(2)}_{zzz}|E_0)\) plane for a TE-polarized plane wave incidence at \(30^\circ\) on a KTP slab. The star corresponds to Fig.~\ref{fig:2HG_field}, where the convergence of the perturbative approach is observed, while the triangle corresponds to Appendix D, Fig.~\ref{fig:2HG_fieldbis} where the approach does not converge.}}
            \label{fig:convergence_radius_2HG}
        \end{figure}

        \smallskip
        \changecolor{The analysis of the convergence radius is inherently complex, as it depends on multiple factors including the geometry of the structure. However, in the case of second-harmonic generation in a slab illuminated by a TE-polarized plane wave, two key parameters are mainly involved in the determination of the radius: the incident-wave amplitude scaled by the susceptibility coefficient, \(|\chi^{(2)}_{zzz}| E_0\), and the ratio of the wavelength \(\lambda_0\) to the slab thickness \(L\). The convergence radius of both the rigorous and perturbative approaches are shown in Fig.~\ref{fig:convergence_radius_2HG}, plotted in the plane \((L/\lambda_0,\ |\chi^{(2)}_{zzz}|E_0)\). The rigorous convergence radius is defined as the maximum incident-wave amplitude for which the Picard iterative algorithm converge, using a tolerance of \(0.001\%\) and a maximum of \(200\) iterations.}
        
        \changecolor{Both convergence radius appear to follow an inverse proportionality law, approximately scaling as \(1/x\), indicating that the thicker the slab, the smaller the incident-wave amplitude must be for methods to converge. The trend is similar for different refractive-index contrasts, as shown in Appendix D, Fig.~\ref{fig:convergence_radius_2HG_indexcontrast}, which can be summarized by the following approximate criterion: \(E_{0, max} \simeq \lambda_0/(L\,\chi^{(2)}_{zzz})\). Moreover, the two radii are close, suggesting a strong correlation between the convergence properties of the two approaches: whenever the rigorous method converges, there is a good chance that the perturbative method also converge. However, this is not always the case, particularly for small slab thicknesses and a very large incident-wave amplitude, as shown in Appendix D, Fig.~\ref{fig:2HG_fieldbis}. The same study was carried out for Kerr nonlinearities, as shown in Appendix D, Fig.~\ref{fig:convergence_radius_kerr}, yielding similar results. 
        }

    \subsection{Computation time}
\begin{figure}
            \centering
            \includegraphics[width=\linewidth]{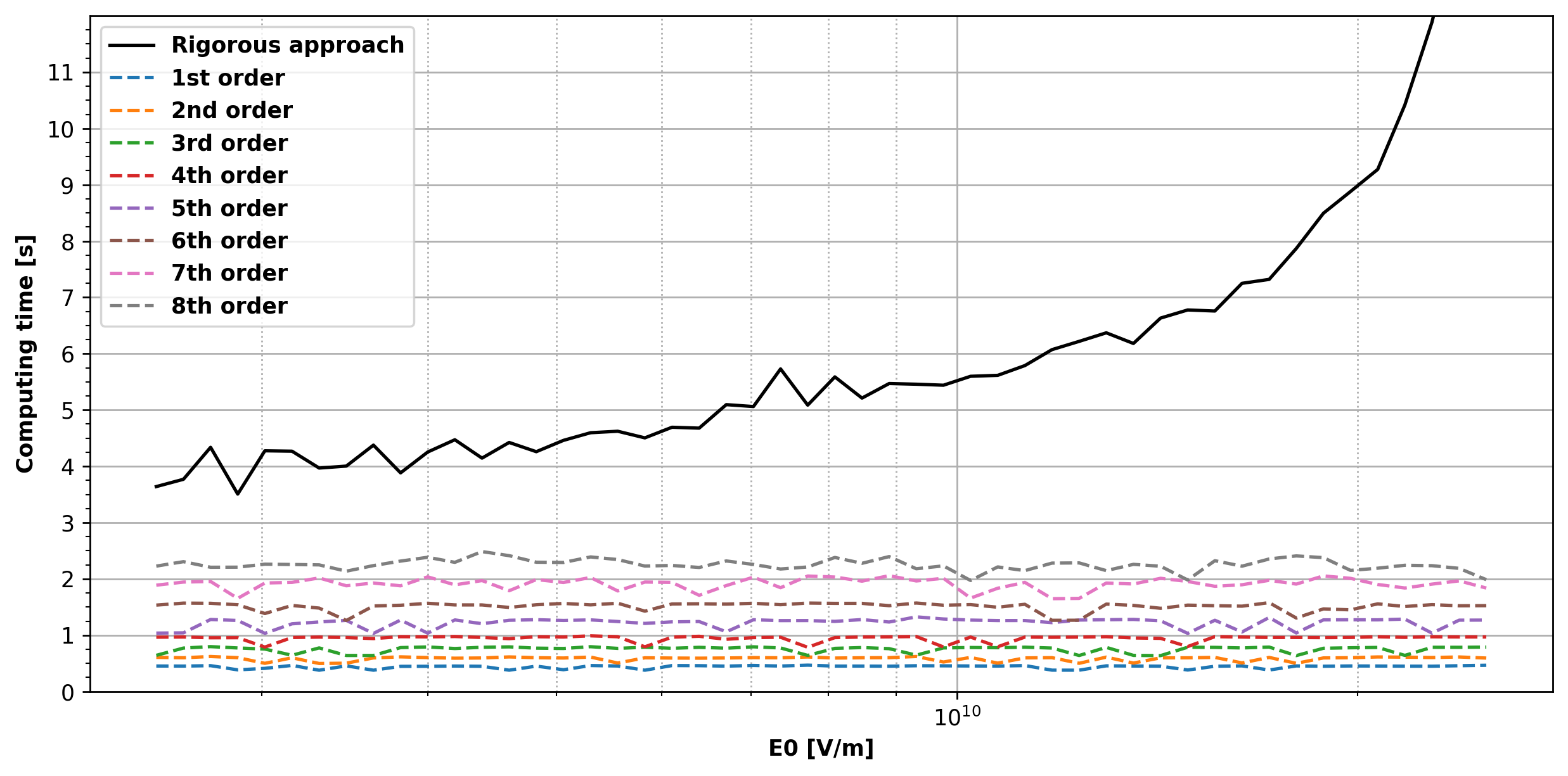}
            \caption{Computation time of the rigorous method and the perturbative one for various expansion orders as a function of the incident-wave amplitude. }
            \label{fig:ComputingTime2HG}
        \end{figure}

        The main advantage of the perturbative approach lies in the significantly reduced computation time required to obtain the solution. It is therefore relevant to compare the computation time between the two approaches for different perturbative orders. Figure~\ref{fig:ComputingTime2HG} shows the simulation time as a function of the incident-wave amplitude, using the same configuration as in Fig.~\ref{fig:2HG_conv}.
        On the one hand, the computation time required by the rigorous approach increases exponentially with the incident-wave amplitude since stronger nonlinear effects demand a greater number of iterations for the simulation to converge. On the other hand, the computation time of the perturbative approach remains essentially constant with respect to the incident-wave amplitude, increasing only gradually with the perturbative order as the number of equations to solve grows.
        Combining the results from Figures~\ref{fig:2HG_conv} and~\ref{fig:ComputingTime2HG}, we observe that using the perturbative approach at high orders enables a drastic reduction in computation time while keeping the error introduced by the approximation relatively low. For instance, for an incident-wave amplitude of \(E_0=3\times10^{9}\,\mathrm{V/m}\), the fourth-order perturbative expansion requires only one quarter of the computation time needed by the rigorous method. At the same amplitude and order, Fig.~\ref{fig:2HG_conv} shows a relative error of about \(e_r\approx 0.1\%\), which may be considered acceptable depending on the intended application.
    
        The method is particularly advantageous for computationally expensive problems, such as those in two or three dimensions. In such cases, the computation time is dominated by the solution of the matrix system generated by the finite element method, while other operations (e.g., assembly and post-processing) become secondary as the problem size increases. Since this step must be repeated at each iteration, the performance gain offered by the perturbative approach can be expected to asymptotically approach the ratio between the number of iterations required by the rigorous method and the number of equations that must be solved in the perturbative approach.
        Furthermore, for large-scale problems, achieving the same level of accuracy as in the 1D case becomes challenging, so the error introduced by the perturbative method will rapidly be overshadowed by the discretization error inherent to the finite element method.
        
\section{Application: Coupled second- and third-order effects}
    \subsection{Governing equations}
        To further illustrate the generality of the proposed method, we examine a scenario incorporating both second- and third-order nonlinearities. By including all the terms of Eq.~\eqref{eq:sys_nl} and truncating the analysis at the third harmonic, we obtain the system given in Sys.~\eqref{eq:3HG}. This system captures several nonlinear effects: second-order processes such as second-harmonic generation \((\omega+\omega\rightarrow2\omega)\) and sum-frequency generation \((2\omega+\omega\rightarrow 3\omega)\); third-harmonic generation \((\omega+\omega+\omega\rightarrow3\omega)\); and both self- and cross-phase modulation resulting from the interactions among different harmonics  \((\omega+q\omega-q\omega\rightarrow\omega)\).
    
        \small
        \begin{equation} 
            \left\{
                \begin{array}{ll}
                    &\mathbf{M}_1^{\mathrm{lin}}\mathbf{E}_1 + \frac{\omega_I^2}{c^2} \Bigl( 2\langle\langle  \mathbf{E}_{-1}, \mathbf{E}_2 \rangle\rangle + 2\langle\langle  \mathbf{E}_{-2}, \mathbf{E}_3 \rangle\rangle \\
                    &+ 3 \langle\langle \mathbf{E}_{-1}, \mathbf{E}_{1}, \mathbf{E}_{1} \rangle\rangle + 6 \langle\langle \mathbf{E}_{-2}, \mathbf{E}_{1}, \mathbf{E}_{2} \rangle\rangle + 6 \langle\langle \mathbf{E}_{-3}, \mathbf{E}_{1}, \mathbf{E}_{3} \rangle\rangle \\
                    &+ 3 \langle\langle \mathbf{E}_{-1}, \mathbf{E}_{-1}, \mathbf{E}_{3} \rangle\rangle + 3 \langle\langle \mathbf{E}_{2}, \mathbf{E}_{2}, \mathbf{E}_{-3} \rangle\rangle \Bigr)\\
                    &= -i\omega\mu_0\,\mathbf{J}_{1} \\
                    &\mathbf{M}_2^{\mathrm{lin}}\mathbf{E}_2 + \frac{(2\omega_I)^2}{c^2} \Bigl( 2\langle\langle  \mathbf{E}_{-1}, \mathbf{E}_3 \rangle\rangle + \langle\langle  \mathbf{E}_{1}, \mathbf{E}_1 \rangle\rangle \\
                    &+ 6 \langle\langle \mathbf{E}_{-1}, \mathbf{E}_{2}, \mathbf{E}_{1} \rangle\rangle + 3 \langle\langle \mathbf{E}_{-2}, \mathbf{E}_{2}, \mathbf{E}_{2} \rangle\rangle + 6 \langle\langle \mathbf{E}_{-3}, \mathbf{E}_{2}, \mathbf{E}_{3} \rangle\rangle \\
                    &+ 6 \langle\langle \mathbf{E}_{-2}, \mathbf{E}_{1}, \mathbf{E}_{3} \rangle\rangle \Bigr)\\
                    &= 0 \\
                    &\mathbf{M}_3^{\mathrm{lin}}\mathbf{E}_3 + \frac{(3\omega_I)^2}{c^2} \Bigl( 2\langle\langle  \mathbf{E}_{1}, \mathbf{E}_2 \rangle\rangle\\
                    &+ 6 \langle\langle \mathbf{E}_{-1}, \mathbf{E}_{3}, \mathbf{E}_{1} \rangle\rangle + 6 \langle\langle \mathbf{E}_{-2}, \mathbf{E}_{3}, \mathbf{E}_{2} \rangle\rangle + 3 \langle\langle \mathbf{E}_{-3}, \mathbf{E}_{2}, \mathbf{E}_{3} \rangle\rangle \\
                    &+ 3 \langle\langle \mathbf{E}_{-1}, \mathbf{E}_{2}, \mathbf{E}_{2} \rangle\rangle + \langle\langle \mathbf{E}_{1}, \mathbf{E}_{1}, \mathbf{E}_{1} \rangle\rangle \Bigr)\\
                    &= 0
                    \label{eq:3HG}
                \end{array}
            \right.
        \end{equation}
        \normalsize
        
        The perturbative equations are retrieved from Sys.~\eqref{eq:sys_prtb_explicit}:
        \begin{subequations}
        \small
            \begin{align}
                &\mathbf{M}_1^{\mathrm{lin}}\mathbf{E}_1^{(1)} = -i \omega \mu_0 \,\mathbf{J}_{p}^{(1)} \\
                &\mathbf{M}_2^{\mathrm{lin}}\mathbf{E}_2^{(2)} + \frac{\left(2\omega_I\right)^2}{c^2} \left( \langle\langle  \mathbf{E}_{1}^{(1)}, \mathbf{E}_{1}^{(1)} \rangle\rangle \right) = \mathbf{0} \\
                &\mathbf{M}_1^{\mathrm{lin}}\mathbf{E}_1^{(3)} + \frac{\omega_I^2}{c^2} \left( 2\langle\langle  \mathbf{E}_{-1}^{(1)}, \mathbf{E}_{2}^{(2)} \rangle\rangle + 3\langle\langle  \mathbf{E}_{1}^{(1)}, \mathbf{E}_{1}^{(1)}, \mathbf{E}_{-1}^{(1)} \rangle\rangle\right) = \mathbf{0} \\
                &\mathbf{M}_3^{\mathrm{lin}}\mathbf{E}_3^{(3)} + \frac{(3\omega_I)^2}{c^2} \left( 2\langle\langle  \mathbf{E}_{1}^{(1)}, \mathbf{E}_{2}^{(2)} \rangle\rangle + \langle\langle  \mathbf{E}_{1}^{(1)}, \mathbf{E}_{1}^{(1)}, \mathbf{E}_{1}^{(1)} \rangle\rangle \right) = \mathbf{0} \\
                &\mathbf{M}_2^{\mathrm{lin}}\mathbf{E}_2^{(4)} + \frac{(2\omega_I)^2}{c^2} \Bigl( 2\langle\langle \mathbf{E}_{1}^{(1)}, \mathbf{E}_{1}^{(3)} \rangle\rangle + 2\langle\langle  \mathbf{E}_{-1}^{(1)}, \mathbf{E}_{3}^{(3)} \rangle\rangle \nonumber \\
                & + 6\langle\langle  \mathbf{E}_{1}^{(1)}, \mathbf{E}_{-1}^{(1)}, \mathbf{E}_{2}^{(2)} \rangle\rangle \Bigr) = \mathbf{0} \\
                &\mathbf{M}_1^{\mathrm{lin}}\mathbf{E}_1^{(5)} + \frac{\omega_I^2}{c^2} \Bigl( 2\langle\langle \mathbf{E}_{2}^{(2)}, \mathbf{E}_{-1}^{(3)} \rangle\rangle + 2\langle\langle  \mathbf{E}_{-1}^{(1)}, \mathbf{E}_{2}^{(4)} \rangle\rangle \nonumber\\
                &+ 2\langle\langle \mathbf{E}_{-2}^{(2)}, \mathbf{E}_{3}^{(3)} \rangle\rangle + 3\langle\langle  \mathbf{E}_{1}^{(1)}, \mathbf{E}_{1}^{(1)}, \mathbf{E}_{-1}^{(3)} \rangle\rangle + 6\langle\langle  \mathbf{E}_{1}^{(1)}, \mathbf{E}_{-1}^{(1)}, \mathbf{E}_{1}^{(3)} \rangle\rangle \nonumber\\
                &+ 3\langle\langle  \mathbf{E}_{-1}^{(1)}, \mathbf{E}_{-1}^{(1)}, \mathbf{E}_{3}^{(3)} \rangle\rangle + 6\langle\langle  \mathbf{E}_{2}^{(2)}, \mathbf{E}_{-2}^{(2)}, \mathbf{E}_{1}^{(1)} \rangle\rangle \Bigr) = \mathbf{0} \\
                &\mathbf{M}_3^{\mathrm{lin}}\mathbf{E}_3^{(5)} + \frac{(3\omega_I)^2}{c^2} \Bigl( 2\langle\langle \mathbf{E}_1^{(1)}, \mathbf{E}_{2}^{(4)} \rangle\rangle + 2\langle\langle  \mathbf{E}_{2}^{(2)}, \mathbf{E}_{1}^{(3)} \rangle\rangle \nonumber\\
                &+ 3\langle\langle  \mathbf{E}_{1}^{(3)}, \mathbf{E}_{1}^{(1)}, \mathbf{E}_{1}^{(1)} \rangle\rangle + 6\langle\langle  \mathbf{E}_{1}^{(1)}, \mathbf{E}_{-1}^{(1)}, \mathbf{E}_{3}^{(3)} \rangle\rangle \nonumber\\
                &+ 3\langle\langle  \mathbf{E}_{2}^{(2)}, \mathbf{E}_{2}^{(2)}, \mathbf{E}_{-1}^{(1)} \rangle\rangle \Bigr) = \mathbf{0}
            \end{align}
            \label{eq:sys_prtb_3HG}
            \normalsize
        \end{subequations}

    \subsection{Numerical results}
\begin{figure*}
            \centering
            \includegraphics[width=0.8\linewidth]{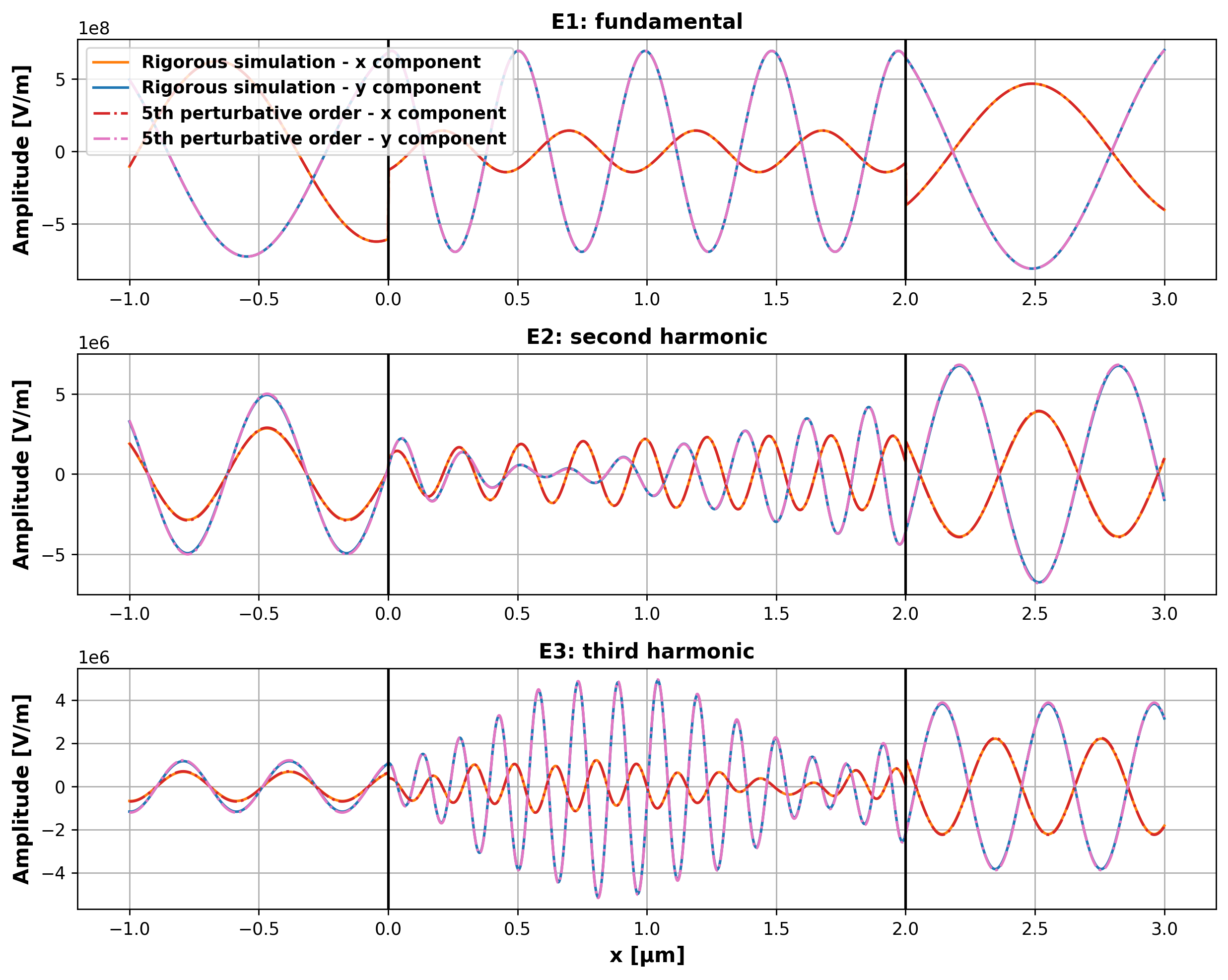}
            \caption{Nonlinear scattering from a KTP slab under TM-polarized illumination. The real parts of the fundamental field \(\mathbf{E}_1\), the second-harmonic \(\mathbf{E}_2\) and the third-harmonic \(\mathbf{E}_3\) are plotted along the \(x\)-axis (\(y=0\)), for an incident plane wave from the left with an amplitude \(A_0=10^{9}V/m\) and an incidence angle \(\theta=30^{\circ}\). The four curves are associated to the rigorous and perturbative methods, for the \(x\)- and \(y\)-components of the fields.} 
            \label{fig:3HG_field}
        \end{figure*}

        Fig.~\ref{fig:3HG_field} shows the real part of the electric field for a TM-polarized plane wave illuminating a nonlinear LiNbO3 slab at an angle \(\theta=30^{\circ}\). The slab is rotated by 90° around the \(y\)-axis so that the wave remains TM-polarized as it propagates through the medium. The linear and nonlinear parameters used in the simulations are listed in Tables~\ref{tab:linear_parameters_LiNbO3}-\ref{tab:nl_parameters_LiNbO3}. The fields are computed using both the rigorous and fifth-order perturbative approaches, with the \(x\)- and \(y\)-components shown. A very good agreement is observed between the two methods, even for the third harmonic, with energy balance errors of \(2\times10^{-6}\%\) and \(6\times10^{-3}\%\), respectively.

        \begin{table}[htbp]
            \centering
            \caption{Linear parameters of LiNbO3 \cite{zelmon_infrared_1997}}
            \begin{tabular}{cc}
                \hline
                \(\lambda \,(\mathrm{nm})\) & \(n_o\) \\
                \hline
                $1064$ & $2.2321$ \\
                $532$ & $2.3232$  \\
                $355$ & $2.4393$  \\
                \hline
            \end{tabular}
            \label{tab:linear_parameters_LiNbO3}
        \end{table}
        \begin{table}[htbp]
            \centering
            \caption{Nonlinear parameters of LiNbO3, at \(\lambda_0 =1064\,\mathrm{nm}\), in [m/V] and [\(\mathrm{m}^2\)/\(\mathrm{V}^2\)] \cite{nikogosyan_nonlinear_2005, kulagin_analysis_2006}}
            \begin{tabular}{ccc}
                \hline
                \(d_{22}\) & \(d_{31}\) & \(d_{33}\) \\
                \hline
                $2.1 \,.10^{-12} $ & $-4.35 \,.10^{-12}$ & $-27.2 \,.10^{-12}$ \\
                \hline
            \end{tabular}
            \begin{tabular}{cccc}
                \hline
                \(\chi_{1111}\) & \(\chi_{1123}\) & \(\chi_{1133}\) & \(\chi_{3333}\) \\
                \hline
                $14.07 \,.10^{-21}$ & $3.35 \,.10^{-21}$ & $3.9 \,.10^{-21}$ & $3.35 \,.10^{-21}$ \\
                \hline
            \end{tabular}
            \label{tab:nl_parameters_LiNbO3}
        \end{table}
        \begin{figure}
            \centering
            \includegraphics[width=\linewidth]{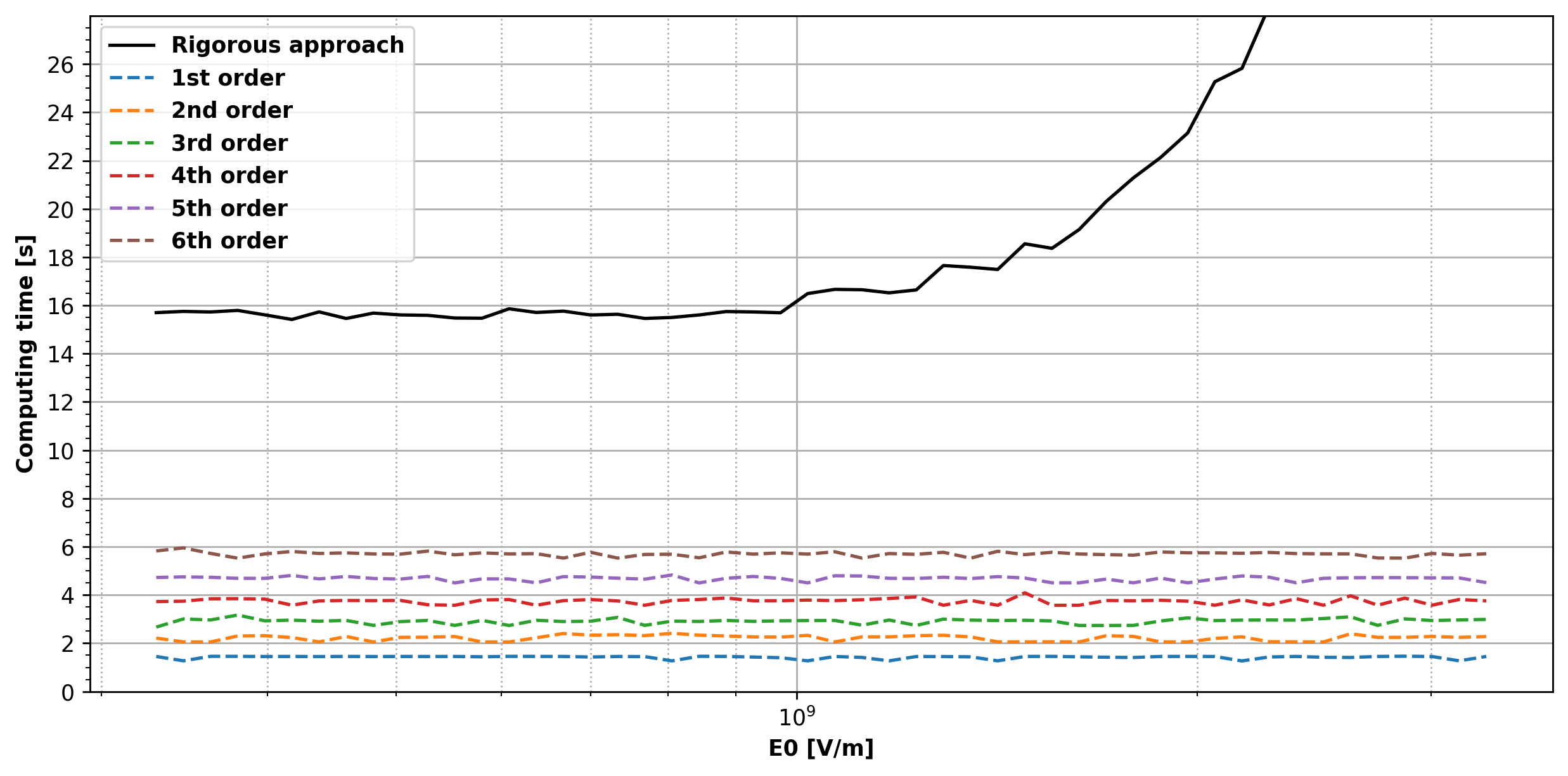}
            \caption{Computation time of the rigorous method and the perturbative one for various expansion orders as a function of the incident-wave amplitude, for a TM-polarized plane wave with an incidence angle \(\theta=30^{\circ}\).}
            \label{fig:ComputingTime3HG}
        \end{figure}
        Fig.~\ref{fig:ComputingTime3HG} shows the computation time for each approach, at six different perturbative orders, for the configuration shown in Fig.~\ref{fig:3HG_field}, as a function of the incident-wave amplitude. As in the 2HG case, we observe a drastic reduction in the computation time when using the perturbative approach. For the incident amplitude used in Fig.~\ref{fig:3HG_field}, namely \(10^9 \,\mathrm{V/m}\), corresponding to the regime just before the sharp increase in computation time observed for the rigorous approach, the fifth-order perturbative method yields a computation time of \(5\,s\) with an energy error of \(6\times10^{-3}\%\), compared with \(16 \,s\) for the rigorous approach.
        Again, we expect the same trend to hold for larger and more complex problems, resulting in a drastic reduction in computation time even at high perturbative orders.

\section{Conclusion and outlook}
    Building on our previous work, we have developed a general approximate method for simulating nonlinear optical interactions with high accuracy, without relying on computationally expensive iterative algorithms. The versatility and performance of the approach are demonstrated through two case studies of light scattering involving two commonly used nonlinear materials, KTP and LiNbO3, with a focus on second-harmonic generation and third-order nonlinear effects. By comparing the method with a more exact iterative scheme previously reported, we quantify the approximation error and identify the range of validity for which the proposed perturbative approach is most advantageous. The substantial reduction in computation time observed in both cases further highlights the benefits of the method. Although the present paper focuses on one-dimensional structures, it should be emphasized that this approach can be applied to more complex, higher-dimensional geometries and source configurations. We further expect that it is precisely in these computationally expensive problems that the method will exhibit its greatest efficiency.
    We hope that this approach will provide a valuable tool for overcoming the limitations of existing models and for investigating more advanced experimental configurations. Owing to its numerical efficiency, it should enable the use of brute-force optimization techniques to design highly optimized nonlinear devices.

\appendix
\section{Appendix: Proofs of the general properties (Sec.3.2)}
    \begin{theorem}
        \(\forall p \in \mathbb{Z},\, \mathbf{E}_{p}\) does not depend on \(\eta\). \label{th:indpt_eta}
    \end{theorem}
    We prove by induction that for all \(j\), the quantity \(\eta^j\mathbf{E}_{p}^{(j)}\) is independent of \(\eta\).
    
    For \(j=1\), we multiply Maxwell's equation by \(\eta\) and substitute the current with its definition:
    \begin{align*}
        \forall p \in \mathbb{Z}, \quad &\mathbf{M}_p^{\mathrm{lin}}\mathbf{E}_p^{(1)} = -i p \omega \mu_0 \,\mathbf{J}_{p}^{(1)} \delta_{|p|,1} \\
        &\mathbf{M}_p^{\mathrm{lin}}\eta\mathbf{E}_p^{(1)} = -i p \omega \mu_0 \,\eta \mathbf{J}_{p}^{(1)} \delta_{|p|,1} \\
        &\mathbf{M}_p^{\mathrm{lin}}\eta\mathbf{E}_p^{(1)} = -i p \omega \mu_0 \, \mathbf{J}_{p} \delta_{|p|,1}
    \end{align*}
    hence, \(\eta\mathbf{E}_p^{(1)}\) does not depend on \(\eta\).
    
    Assuming the property holds for all \(j < j_0\), we now show that it remains valid at order \(j=j_0\), proceeding as before:
    \begin{align}
        &\forall p \in \mathbb{Z},\nonumber\\ &\mathbf{M}_p^{\mathrm{lin}}\mathbf{E}_p^{(j_0)} + \frac{\left(p\omega_I\right)^2}{c^2} \sum_{l+k=j_0} \sum_{q\in \mathbb{Z}} \langle\langle  \mathbf{E}_{q}^{(l)}, \mathbf{E}_{p-q}^{(k)} \rangle\rangle \nonumber\\[-5pt]
        &+ \frac{\left(p\omega_I\right)^2}{c^2} \sum_{l+m+k=j_0} \sum_{(q,r)\in \mathbb{Z}^2} \langle\langle  \mathbf{E}_{q}^{(l)}, \mathbf{E}_{r}^{(m)}, \mathbf{E}_{p-q-r}^{(k)} \rangle\rangle = \mathbf{0} \\[10pt]
        &\mathbf{M}_p^{\mathrm{lin}}\eta^{j_0}\mathbf{E}_p^{(j_0)} + \frac{\left(p\omega_I\right)^2}{c^2} \sum_{l+k=j_0} \sum_{q\in \mathbb{Z}} \eta^{j_0} \langle\langle  \mathbf{E}_{q}^{(l)}, \mathbf{E}_{p-q}^{(k)} \rangle\rangle \nonumber\\[-5pt]
        &+ \frac{\left(p\omega_I\right)^2}{c^2} \sum_{l+m+k=j_0} \sum_{(q,r)\in \mathbb{Z}^2} \eta^{j_0}\langle\langle  \mathbf{E}_{q}^{(l)}, \mathbf{E}_{r}^{(m)}, \mathbf{E}_{p-q-r}^{(k)} \rangle\rangle = \mathbf{0} \\[0pt]
        &\mathbf{M}_p^{\mathrm{lin}}\eta^{j_0}\mathbf{E}_p^{(j_0)} + \frac{\left(p\omega_I\right)^2}{c^2} \sum_{l+k=j_0} \sum_{q\in \mathbb{Z}} \langle\langle \eta^{l} \mathbf{E}_{q}^{(l)}, \eta^{k} \mathbf{E}_{p-q}^{(k)} \rangle\rangle \nonumber\\[-5pt]
        &+ \frac{\left(p\omega_I\right)^2}{c^2} \sum_{l+m+k=j_0} \sum_{(q,r)\in \mathbb{Z}^2} \langle\langle  \eta^l \mathbf{E}_{q}^{(l)}, \eta^m\mathbf{E}_{r}^{(m)}, \eta^k\mathbf{E}_{p-q-r}^{(k)} \rangle\rangle = \mathbf{0} 
    \end{align}
    Based on the induction hypothesis, since \(l<j_0\) and \(k<j_0\) in the second-order term of the equation, each quantity \(\eta^{l} \mathbf{E}_{q}^{(l)}\) and \(\eta^{k} \mathbf{E}_{p-q}^{(k)}\) is independent of \(\eta\). The same argument applies to third-order terms. Therefore, by induction, \(\eta^{j_0} \mathbf{E}_{p}^{(j_0)}\) is also independent of \(\eta\). Hence, \(\mathbf{E}_p=\sum_j \mathbf{E}_p^{(j)}\) does not depend on \(\eta\).

    \bigskip
    
    \begin{theorem}
        \(\forall (p,j) \in \mathbb{Z}\times \mathbb{N}^*,\, |p| > j \implies \mathbf{E}_p^{(j)}=\mathbf{0}\) \label{th:p_higher_j}
    \end{theorem}
    
    We only need to prove the proposition for \(p>0\), since \(E_{-p}^{(j)}=\overline{E_p^{(j)}}\). We proceed by induction on the perturbative order \(j\).
    
    For \(j=1\), the property follows directly from the linear Maxwell equation:
    \begin{equation}
        \forall p \in \mathbb{Z}, \quad \mathbf{M}_p^{\mathrm{lin}}\mathbf{E}_p^{(1)} = -i p \omega \mu_0 \,\mathbf{J}_{p}^{(1)} \delta_{|p|,1}
    \end{equation}
    Then for all \(p > 1\), \(\mathbf{M}_p^{\mathrm{lin}}\mathbf{E}_p^{(1)} = \mathbf{0}\) and thus \(\mathbf{E}_p^{(1)} = \mathbf{0}\).
    
    If the property holds for all \(j < j_0\), then at order \(j=j_0\) we have:
    \begin{align}
    \forall p \in \mathbb{Z}, \quad \mathbf{M}_p^{\mathrm{lin}}\mathbf{E}_p^{(j_0)} + \frac{\left(p\omega_I\right)^2}{c^2} \sum_{l+k=j_0} \sum_{q\in \mathbb{Z}} \langle\langle  \mathbf{E}_{q}^{(l)}, \mathbf{E}_{p-q}^{(k)} \rangle\rangle \nonumber\\[-5pt]+ \frac{\left(p\omega_I\right)^2}{c^2} \sum_{l+m+k=j_0} \sum_{(q,r)\in \mathbb{Z}^2} \langle\langle  \mathbf{E}_{q}^{(l)}, \mathbf{E}_{r}^{(m)}, \mathbf{E}_{p-q-r}^{(k)} \rangle\rangle = \mathbf{0}
     \end{align}
    We first show that for all \(p > j_0\), all \(q \in \mathbb{Z}\) and all \((l,k)\in\mathbb{N}^{*2}\) such that \(l+k=j_0\),
    \begin{equation}
        \langle\langle  \mathbf{E}_{q}^{(l)}, \mathbf{E}_{p-q}^{(k)} \rangle\rangle = 0
    \end{equation}
    Indeed:
    \begin{itemize}
        \item if \(q > l\), then by the induction hypothesis, \(\mathbf{E}_q^{(l)} = \mathbf{0}\)
        \item if \(q \leq l\) then \(-q\geq-l\), and since \(p>j_0\), we have \(p-q > j_0-l = k\). Therefore, \(\mathbf{E}_{p-q}^{(k)} = \mathbf{0}\).
    \end{itemize}
    We now show that for all \(p > j_0\), all \((q,r) \in \mathbb{Z}^2\), and all \((l,m,k)\in\mathbb{N}^2\) such that \(l+m+k=j_0\),
    \begin{equation}
        \langle\langle  \mathbf{E}_{q}^{(l)}, \mathbf{E}_{r}^{(m)}, \mathbf{E}_{p-q-r}^{(k)} \rangle\rangle = 0
    \end{equation}
    Indeed:
    \begin{itemize}
        \item if \(q > l\) or \(r > m\), then by the induction hypothesis, the corresponding fields vanish, and thus \(\langle\langle  \mathbf{E}_{q}^{(l)}, \mathbf{E}_{r}^{(m)}, \mathbf{E}_{p-q-r}^{(k)} \rangle\rangle = 0\)
        \item if \(q \leq l\) and \(r \leq m\), then \(-q\geq-l\) and \(-r \geq -m\). Since \(p>j_0\), it follows that \(p-q-r > j_0-l-m = k\), hence \(\mathbf{E}_{p-q-r}^{(k)} = \mathbf{0}\).
    \end{itemize}
    Therefore, in all cases,
    \begin{equation}
        \langle\langle  \mathbf{E}_{q}^{(l)}, \mathbf{E}_{p-q}^{(k)} \rangle\rangle = \langle\langle  \mathbf{E}_{q}^{(l)}, \mathbf{E}_{r}^{(m)}, \mathbf{E}_{p-q-r}^{(k)} \rangle\rangle = \mathbf{0}
    \end{equation}
    which implies that \(\mathbf{E}_{p}^{(j_0)} = \mathbf{0}\).
    
    \bigskip
    \begin{theorem}
        \(\forall p \in \mathbb{Z},\, p-j = 1[2] \implies \mathbf{E}_p^{(j)}=\mathbf{0}\)
        
        In other words, if \(p\) and \(j\) don't have the same parity, then \(\mathbf{E}_p^{(j)}=\mathbf{0}\). \label{th:parity}
    \end{theorem}
    
    Again, we proceed by induction on the perturbative order \(j\).
    For \(j=1\), only the term \(\mathbf{E}_1^{(1)}\) is nonzero, while all other fields vanish. Hence, the proposition holds.
    
    Assume now that the property holds for all \(j < j_0\). Then at order \(j=j_0\), we have:
    \begin{align}
        \forall p \in \mathbb{Z}, \quad \mathbf{M}_p^{\mathrm{lin}}\mathbf{E}_p^{(j_0)} + \frac{\left(p\omega_I\right)^2}{c^2} \sum_{l+k=j_0} \sum_{q\in \mathbb{Z}} \langle\langle  \mathbf{E}_{q}^{(l)}, \mathbf{E}_{p-q}^{(k)} \rangle\rangle \nonumber\\[-5pt]+ \frac{\left(p\omega_I\right)^2}{c^2} \sum_{l+m+k=j_0} \sum_{(q,r)\in \mathbb{Z}^2} \langle\langle  \mathbf{E}_{q}^{(l)}, \mathbf{E}_{r}^{(m)}, \mathbf{E}_{p-q-r}^{(k)} \rangle\rangle = \mathbf{0}
     \end{align}
    We first show that for all \(p \in \mathbb{Z}\) such that \(p-j_0=1[2]\), all \(q \in \mathbb{Z}\) and all \((l,k)\in\mathbb{N}^2\) such that \(l+k=j_0\),
    \begin{equation}
    \langle\langle \mathbf{E}_{q}^{(l)}, \mathbf{E}_{p-q}^{(k)} \rangle\rangle = 0
    \end{equation}
    
    Let us consider the last term \(\mathbf{E}_{p-q}^{(k)}\):
    \begin{align*}
        (p-q)-k &= (p-q) - (j_0 - l) \\
        &= \underbrace{(p - j_0)}_{=1[2]} - (q-l)
    \end{align*}
    
    If (\(q-l\)) is odd, then (\((p-q)-k\)) is even, and vice versa. Therefore, in all cases, one of the two fields vanishes according to the induction hypothesis, which implies \(\langle\langle  \mathbf{E}_{q}^{(l)}, \mathbf{E}_{p-q}^{(k)} \rangle\rangle = 0\).
    
    Next, we show that for all \(p \in \mathbb{Z}\) such that \(p-j_0=1[2]\), all \((q,r) \in \mathbb{Z}^2\), and all \((l,m,k)\in\mathbb{N}^3\) such that \(l+m+k=j_0\),
    \begin{equation}
    \langle\langle \mathbf{E}_{q}^{(l)}, \mathbf{E}_{r}^{(m)}, \mathbf{E}_{p-q-r}^{(k)} \rangle\rangle = 0
    \end{equation}
    
    Considering the last term \(\mathbf{E}_{p-q-r}^{(k)}\):
    \begin{align*}
        (p-q-r)-k &= (p-q-r) - (j_0 - l-m) \\
        &= \underbrace{(p - j_0)}_{=1[2]} - (q-l) - (r-m)
    \end{align*}
    
    If either (\(q-l\)) or (\(r-m\)) is odd, then \(\langle\langle  \mathbf{E}_{q}^{(l)}, \mathbf{E}_{r}^{(m)}, \mathbf{E}_{p-q-r}^{(k)} \rangle\rangle=0\). If both are even, then \(-(q-l)-(r-m)=0[2]\) and \((p-q-r)-k=1[2]\), again implying that the corresponding field vanishes.
    
    Hence, in all cases,
    \begin{equation}
        \langle\langle  \mathbf{E}_{q}^{(l)}, \mathbf{E}_{p-q}^{(k)} \rangle\rangle = \langle\langle  \mathbf{E}_{q}^{(l)}, \mathbf{E}_{r}^{(m)}, \mathbf{E}_{p-q-r}^{(k)} \rangle\rangle = \mathbf{0}
    \end{equation}
    and therefore \(\mathbf{E}_{p}^{(j_0)} = \mathbf{0}\).


\section{Appendix: Nonlinear sources}
    The nonlinear sources take the following form:
    \begin{align}
        \mathbf{S}_p^{(j)} = \frac{(p\omega_I)^2}{c^2} \Bigl( \sum_{k+l=j} \sum_{q\in\mathbb{Z}}\langle\langle  \mathbf{E}_{q}^{(k)}, \mathbf{E}_{p-q}^{(l)} \rangle\rangle \nonumber\\
        + \sum_{k+l+m=j} \sum_{(q,r)\in\mathbb{Z}^{2}} \langle\langle \mathbf{E}_{q}^{(k)}, \mathbf{E}_{r}^{(l)}, \mathbf{E}_{p-q-r}^{(m)} \rangle\rangle \Bigr)
    \end{align}

    By using properties (i) and (ii) from Sec.~3, we drastically reduce the number of terms in the summation. The remaining terms for the first five perturbative orders are:
    \begin{subequations}
        \begin{align}
            \mathbf{S}_2^{(2)} &= \frac{\left(2\omega_I\right)^2}{c^2} \langle\langle  \mathbf{E}_{1}^{(1)}, \mathbf{E}_{1}^{(1)} \rangle\rangle \\
            \mathbf{S}_1^{(3)} &= \frac{\omega_I^2}{c^2} \left( 2\langle\langle  \mathbf{E}_{-1}^{(1)}, \mathbf{E}_{2}^{(2)} \rangle\rangle + 3\langle\langle  \mathbf{E}_{1}^{(1)}, \mathbf{E}_{1}^{(1)}, \mathbf{E}_{-1}^{(1)} \rangle\rangle \right) \\
            \mathbf{S}_3^{(3)} &= \frac{(3\omega_I)^2}{c^2} \left( 2\langle\langle  \mathbf{E}_{1}^{(1)}, \mathbf{E}_{2}^{(2)} \rangle\rangle + \langle\langle  \mathbf{E}_{1}^{(1)}, \mathbf{E}_{1}^{(1)}, \mathbf{E}_{1}^{(1)} \rangle\rangle \right) \\
            \mathbf{S}_2^{(4)} &= \frac{(2\omega_I)^2}{c^2} \Bigl( 2\langle\langle \mathbf{E}_{1}^{(1)}, \mathbf{E}_{1}^{(3)} \rangle\rangle + 2\langle\langle  \mathbf{E}_{-1}^{(1)}, \mathbf{E}_{3}^{(3)} \rangle\rangle \nonumber\\
            & + 6\langle\langle  \mathbf{E}_{1}^{(1)}, \mathbf{E}_{-1}^{(1)}, \mathbf{E}_{2}^{(2)} \rangle\rangle \Bigr) \\
            \mathbf{S}_4^{(4)} &= \frac{(4\omega_I)^2}{c^2} \Bigl( 2\langle\langle \mathbf{E}_{1}^{(1)}, \mathbf{E}_{3}^{(3)} \rangle\rangle + \langle\langle  \mathbf{E}_{2}^{(2)}, \mathbf{E}_{2}^{(2)} \rangle\rangle \nonumber\\
            & + 3\langle\langle  \mathbf{E}_{2}^{(1)}, \mathbf{E}_{1}^{(1)}, \mathbf{E}_{1}^{(1)} \rangle\rangle \Bigr) \\
            \mathbf{S}_1^{(5)} &= \frac{\omega_I^2}{c^2} \Bigl( 2\langle\langle \mathbf{E}_{2}^{(2)}, \mathbf{E}_{-1}^{(3)} \rangle\rangle + 2\langle\langle  \mathbf{E}_{-1}^{(1)}, \mathbf{E}_{2}^{(4)} \rangle\rangle \nonumber\\
            &+ 2\langle\langle \mathbf{E}_{-2}^{(2)}, \mathbf{E}_{3}^{(3)} \rangle\rangle + 3\langle\langle  \mathbf{E}_{1}^{(1)}, \mathbf{E}_{1}^{(1)}, \mathbf{E}_{-1}^{(3)} \rangle\rangle \nonumber\\
            &+ 6\langle\langle  \mathbf{E}_{1}^{(1)}, \mathbf{E}_{-1}^{(1)}, \mathbf{E}_{1}^{(3)} \rangle\rangle + 3\langle\langle  \mathbf{E}_{-1}^{(1)}, \mathbf{E}_{-1}^{(1)}, \mathbf{E}_{3}^{(3)} \rangle\rangle \nonumber\\
            &+ 6\langle\langle  \mathbf{E}_{2}^{(2)}, \mathbf{E}_{-2}^{(2)}, \mathbf{E}_{1}^{(1)} \rangle\rangle \Bigr) \\
            \mathbf{S}_3^{(5)} &= \frac{(3\omega_I)^2}{c^2} \Bigl( 2\langle\langle \mathbf{E}_1^{(1)}, \mathbf{E}_{2}^{(4)} \rangle\rangle + 2\langle\langle  \mathbf{E}_{2}^{(2)}, \mathbf{E}_{1}^{(3)} \rangle\rangle \nonumber\\
            &+ 2\langle\langle \mathbf{E}_{-1}^{(1)}, \mathbf{E}_{4}^{(4)} \rangle\rangle  + 3\langle\langle  \mathbf{E}_{1}^{(3)}, \mathbf{E}_{1}^{(1)}, \mathbf{E}_{1}^{(1)} \rangle\rangle  \nonumber\\
            & 6\langle\langle  \mathbf{E}_{1}^{(1)}, \mathbf{E}_{-1}^{(1)}, \mathbf{E}_{3}^{(3)} \rangle\rangle + 3\langle\langle  \mathbf{E}_{2}^{(2)}, \mathbf{E}_{2}^{(2)}, \mathbf{E}_{-1}^{(1)} \rangle\rangle \Bigr) \\
            \mathbf{S}_5^{(5)} &= \frac{(5\omega_I)^2}{c^2} \Bigl( 2\langle\langle \mathbf{E}_{1}^{(1)}, \mathbf{E}_{4}^{(4)} \rangle\rangle + 2\langle\langle  \mathbf{E}_{2}^{(2)}, \mathbf{E}_{3}^{(3)} \rangle\rangle \nonumber\\
            &+ 3\langle\langle  \mathbf{E}_{3}^{(3)}, \mathbf{E}_{1}^{(1)}, \mathbf{E}_{1}^{(1)} \rangle\rangle + 3\langle\langle  \mathbf{E}_{2}^{(2)}, \mathbf{E}_{2}^{(2)}, \mathbf{E}_{1}^{(1)} \rangle\rangle \Bigr)
        \end{align}
        \label{eq:sys_prtb_explicit_spj}
    \end{subequations}

\section{Appendix: 2D to 1D problem}
    It is shown in \cite{itier_scattering_2025} that the problem reduces to a one-dimensional formulation by replacing the linear operator \(\mathbf{M}^{\mathrm{\mathrm{lin}}}_p\) by \(\mathbf{\tilde{M}}^{\mathrm{lin}}_p\), defined as:
        \begin{equation}
           \mathbf{\tilde{M}}^{\mathrm{lin}}_p\left(\mathbf{\tilde{E}}_p^j(x)\right) \equiv  \mathbf{M}^{\mathrm{lin}}_p\left(\mathbf{\tilde{E}}_p^j(x) e^{ip\beta y}\right) e^{-ip\beta y}  
        \end{equation}
    The operator is explicitly defined below:
    \begin{align}
        \mathbf{\tilde{M}}_p^{\mathrm{lin}}(\mathbf{\tilde{E}}_p) = -\mathbf{T}_p(\mathbf{\tilde{E}}_p) + \frac{\left(p\omega_I\right)^2}{c^2} \varepsilon_r \cdot \mathbf{\tilde{E}}_p
    \end{align}

    By denoting \(\mathbf{u}=ip\beta\,\mathbf{e}_y\), with \(\beta=k_0\sin{\theta}\), the operator \(\mathbf{T}_p\) can be written as:
    \begin{align}
        \mathbf{T}_p(\mathbf{\tilde{E}}_p) =& \Bigr( \nabla \times \nabla \times \mathbf{\tilde{E}}_p + \nabla \times \left(\mathbf{u} \times \mathbf{\tilde{E}}_p\right) \nonumber\\
        &+ \mathbf{u} \times \left( \nabla \times \mathbf{\tilde{E}}_p\right) + \mathbf{u} \times (\mathbf{u} \times \mathbf{\tilde{E}}_p) \Bigl) \nonumber\\
        =&
        \begin{pmatrix}
                                & \; &ip\beta \partial_x (\tilde{E}_p)_y   & + & (p\beta)^2 (\tilde{E}_p)_x \\
            -\partial^2_x (\tilde{E}_p)_y & + &ip\beta \partial_x (\tilde{E}_p)_x   &   & \\
            -\partial^2_x (\tilde{E}_p)_z &   &                           & + & (p\beta)^2 (\tilde{E}_p)_z
        \end{pmatrix} \label{eq:Tp2}
    \end{align}
    With \((\tilde{E}_p)_i\) being the i-th component of the field \(\tilde{E}_p\).

\section{Appendix: Complementary results on the convergence radius of the perturbative approach}
\begin{figure}
        \centering
        \includegraphics[width=\linewidth]{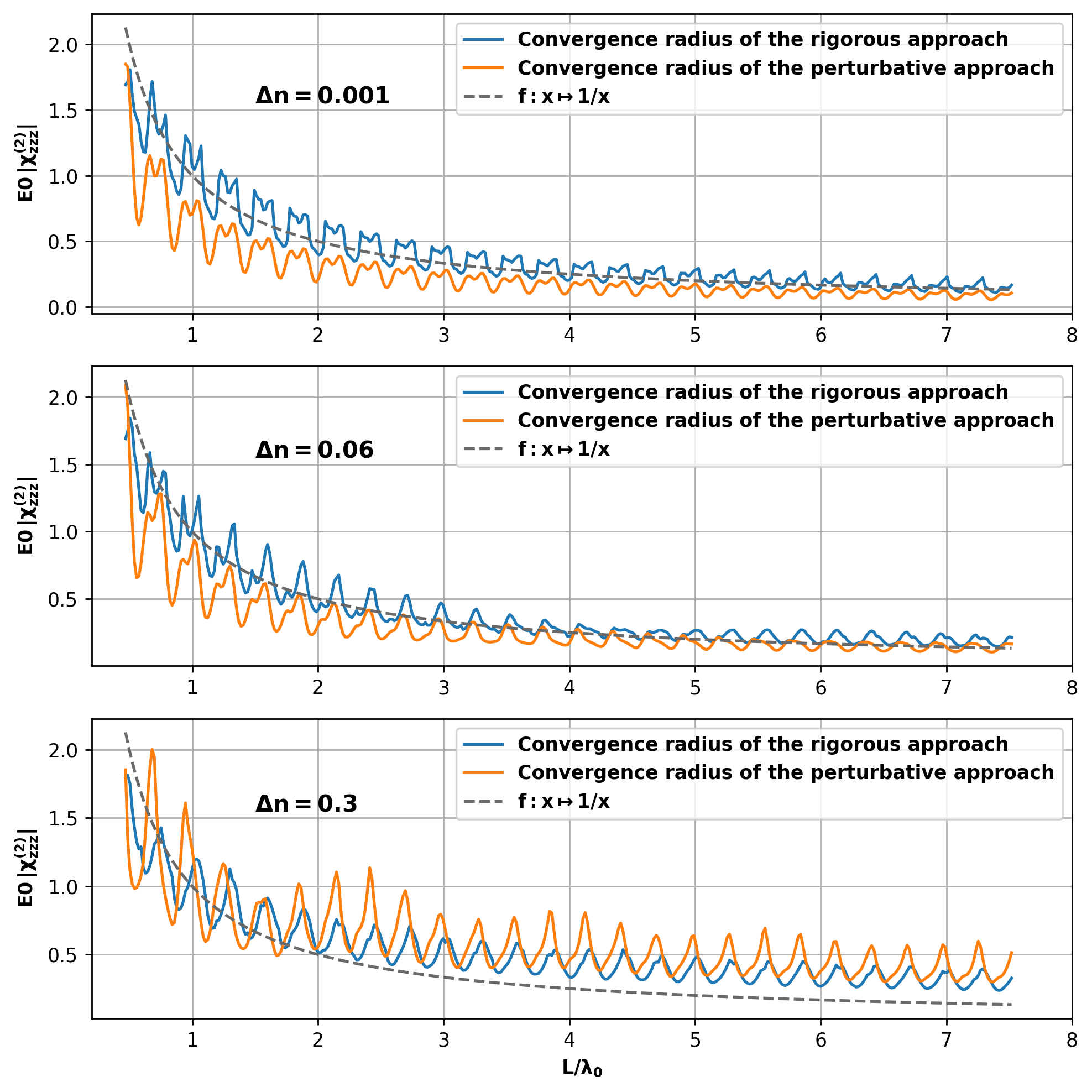}
        \caption{\changecolor{Convergence radius plotted in the \((L/\lambda_0,\ |\chi^{(2)}_{zzz}|E_0)\) plane for a TE-polarized plane wave incidence at \(30^\circ\), when considering second-harmonic generation only. Three different refractive-index contrasts are analyzed: \(0.001, \ 0.06, \ 0.3\).}}
        \label{fig:convergence_radius_2HG_indexcontrast}
    \end{figure}

\begin{figure*}
        \centering
        \includegraphics[width=0.8\linewidth]{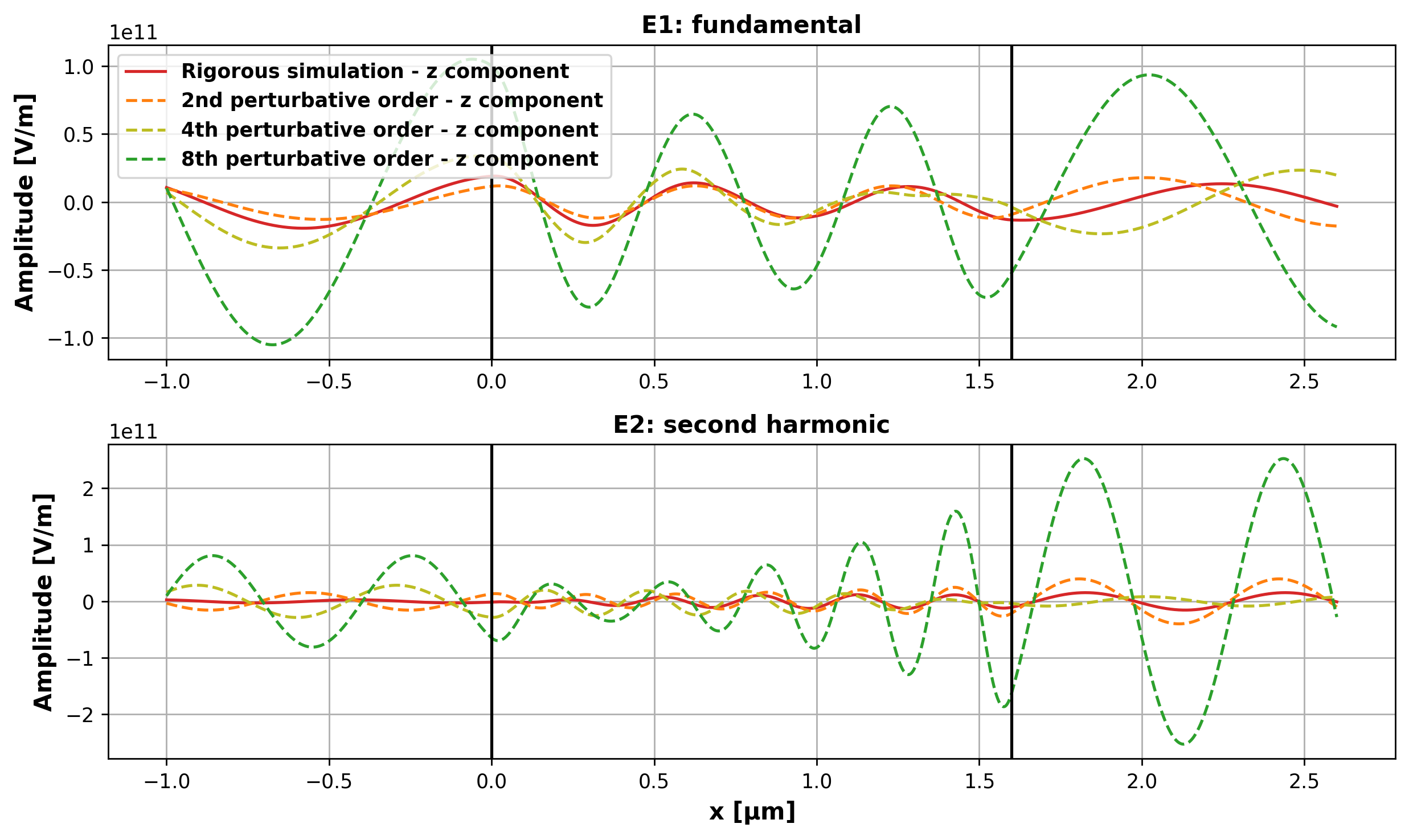}
        \caption{\changecolor{Example of a non-convergent case. The figure shows nonlinear scattering from a KTP slab under TE-polarized illumination. The material is the same as Fig.~\ref{fig:2HG_field} but the width of the slab is different (\(L=1.6\,\mathrm{\mu m}\)). The real parts of the fundamental field \(\mathbf{E}_1\) and the second-harmonic \(\mathbf{E}_2\) are plotted along the \(x\)-axis (\(y=0\)), for an incident plane wave from the left with amplitude \(A_0=2.1\times10^{10}V/m\) and incidence angle \(\theta=30^{\circ}\). The four curves correspond to the rigorous method and to the perturbative approach at second, fourth, and eighth order, respectively.}} 
        \label{fig:2HG_fieldbis}
    \end{figure*}

\begin{figure}
        \centering
        \includegraphics[width=\linewidth]{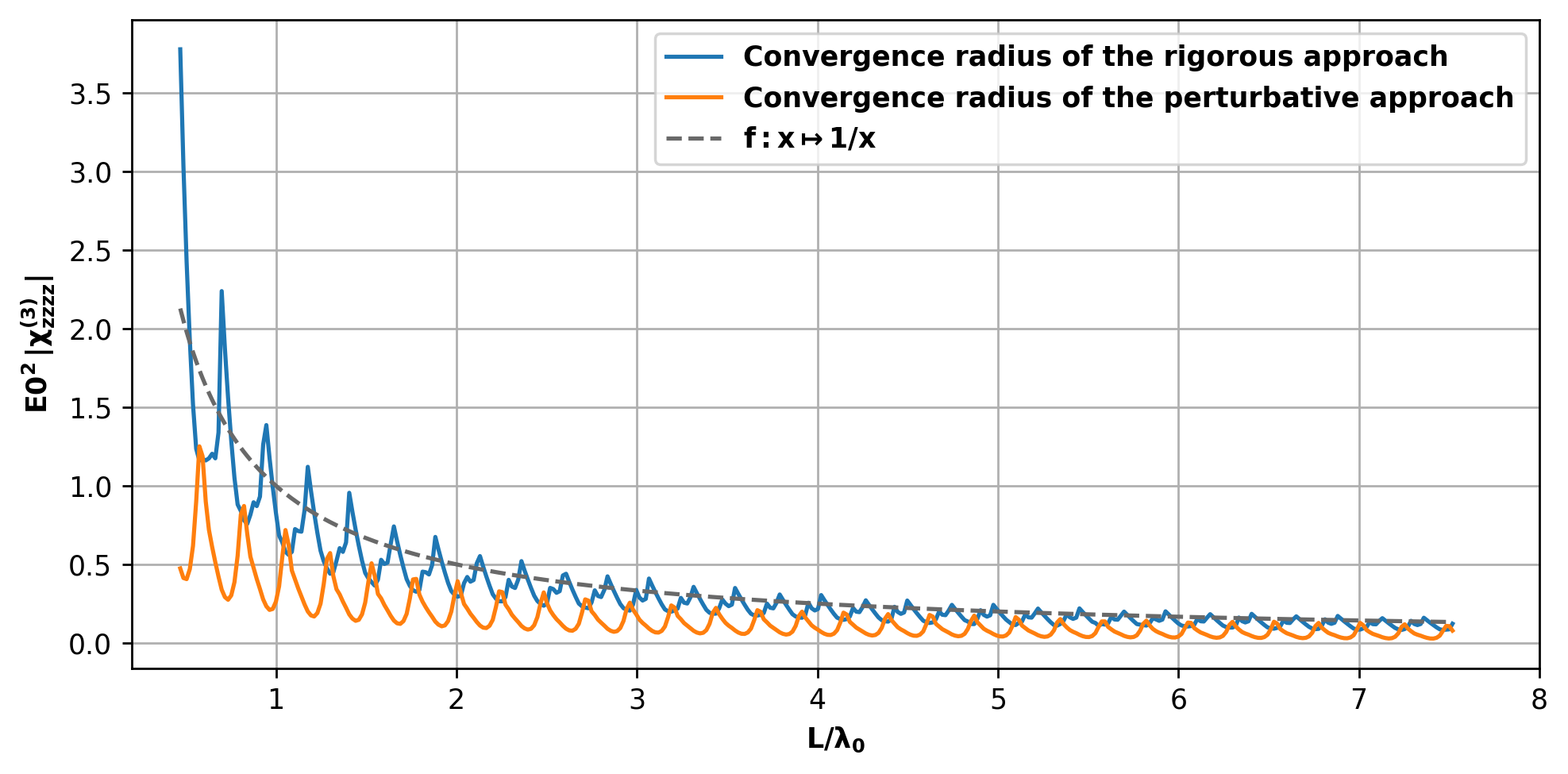}
        \caption{\changecolor{Convergence radius plotted in the \((L/\lambda_0,\ |\chi^{(3)}_{zzzz}|(E_0)^2)\) plane for a TE-polarized plane wave incidence at \(30^\circ\), when considering Kerr effect only. The material used is LiNbO3, the same as Fig.~\ref{fig:3HG_field}, described in Tables~\ref{tab:linear_parameters_LiNbO3} and \ref{tab:nl_parameters_LiNbO3}}.}
        \label{fig:convergence_radius_kerr}
    \end{figure}

    \changecolor{We show below additional studies on the convergence behavior of the rigorous and perturbative approaches. Fig.~\ref{fig:convergence_radius_2HG_indexcontrast} shows the convergence radius when only second-harmonic generation is considered, plotted for three different refractive index contrasts \(0.001, \ 0.06, \ 0.3\). In this case, the convergence radius can be approximated by \(E_{0, max} \simeq \lambda_0/(L\,\chi^{(2)}_{zzz})\). Fig.~\ref{fig:2HG_fieldbis} illustrates a case in which the perturbative approach fails to converge. Fig.~\ref{fig:convergence_radius_kerr} shows the convergence radius when only Kerr effect is considered. In this case, the radius can be approximated by \(E_{0, max} \simeq \sqrt{\lambda_0/(L\,\chi^{(3)}_{zzzz})}\). }
    
\printcredits

\bibliographystyle{cas-model2-names}

\bibliography{cas-refs}


\end{document}